\documentclass[fleqn,usenatbib,useAMS]{mnras}

\usepackage{graphicx}
\usepackage{amsmath}
\usepackage{amssymb}
\usepackage{multicol}      
\usepackage{bm}	
\usepackage{threeparttable}
\usepackage{tabularray}
\usepackage{longtable}
\usepackage{multirow}
\usepackage{booktabs}
\usepackage{lscape}
\usepackage{rotating}
\usepackage{adjustbox}

\newcommand{\kms}{\,km\,s$^{-1}$}

\usepackage[T1]{fontenc}

\usepackage{newtxtext,newtxmath}

\title[Chemical abundances of the open cluster NGC 6705]{Chemical abundances of the young inner-disk open cluster NGC 6705 observed by APOGEE: sodium-rich and not $\alpha$-enhanced}

\author[Loaiza-Tacuri et al.]{
V. Loaiza-Tacuri,$^{1}$\thanks{Contact e-mail: \href{mailto:vtacuri@on.br}{vtacuri@on.br}}
K. Cunha,$^{1,2,3}$
D. Souto,$^{4}$
V. V. Smith,$^{5,3}$
R. Guerço,$^{1}$
C. Chiappini,$^{6}$
J. V. Sales Silva,$^{1}$
\newauthor
D. Horta,$^{7}$
C. Allende Prieto,$^{8}$
R. Beaton,$^{9}$
D. Bizyaev,$^{10,11}$
S. Daflon,$^{1}$
P. Frinchaboy,$^{12}$
S. Hasselquist,$^{13}$
\newauthor
C. R. Hayes,$^{14}$
J. A. Holtzman,$^{15}$
H. J\"onsson,$^{16}$
S. R. Majewski,$^{17}$
S. M{\'e}sz{\'a}ros,$^{18,19}$
\newauthor
D. L. Nidever,$^{20}$
M. Pinsonneault,$^{21}$
G. Zasowski,$^{22}$
\\
$^{1}$Observat\'orio Nacional, Rua General Jos\'e Cristino, 77, 20921-400 S\~ao Crist\'ov\~ao, Rio de Janeiro, RJ, Brazil\\
$^{2}$Steward Observatory, University of Arizona, 933 North Cherry Avenue, Tucson, AZ 85721-0065, USA\\
$^{3}$Institut d'Astrophysique de Paris, UMR7095 CNRS, Sorbonne Universit\'e, 98bis Bd. Arago, 75014 Paris, France\\
$^{4}$Departamento de F\'isica, Universidade Federal de Sergipe, Av. Marechal Rondon, S/N, 49000-000 S\~ao Crist\'ov\~ao, SE, Brazil\\
$^{5}$NSF’s NOIRLab, 950 N. Cherry Ave. Tucson, AZ 85719 USA\\
$^{6}$Leibniz-Institut f\"ur Astrophysik Potsdam (AIP), An der Sternwarte 16, 14482 Potsdam, Germany \\
$^{7}$Center for Computational Astrophysics, Flatiron Institute,162 Fifth Avenue, New York, NY 10010, USA \\
$^{8}$Departamento de Astrofísica, Universidad de La Laguna, E-38206 La Laguna, Tenerife, Spain \\
$^{9}$The Observatories of the Carnegie Institution for Science, 813 Santa Barbara Street, Pasadena, CA 91101, USA \\
$^{10}$Apache Point Observatory and New Mexico State University, P.O. Box 59, Sunspot, NM 88349-0059, USA \\
$^{11}$Sternberg Astronomical Institute, Moscow State University, Moscow, Russia \\
$^{12}$Department of Physics \& Astronomy, Texas Christian University, TCU Box 298840, Fort Worth, TX 76129, USA\\
$^{13}$Space Telescope Science Institute, 3700 San Martin Drive, Baltimore, MD 21218, USA \\
$^{14}$NRC Herzberg Astronomy and Astrophysics Research Centre, 5071 West Saanich Road, Victoria, B.C., Canada, V9E 2E7 \\
$^{15}$ New Mexico State University, Las Cruces, NM 88003, USA\\
$^{16}$Materials Science and Applied Mathematics, Malm\"o University, SE-205 06 Malm\"o, Sweden \\
$^{17}$Department of Astronomy, University of Virginia, Charlottesville, VA 22904-4325, USA \\
$^{18}$ELTE E\"otv\"os Lor\'and University, Gothard Astrophysical Observatory, 9700 Szombathely, Szent Imre H. st. 112, Hungary \\
$^{19}$MTA-ELTE Lend{\"u}let "Momentum" Milky Way Research Group, Hungary \\
$^{20}$Department of Physics, Montana State University, P.O. Box 173840, Bozeman, MT 59717-3840 \\
$^{21}$Department of Astronomy, The Ohio State University, Columbus, OH 43210, USA \\
$^{22}$Department of Physics and Astronomy, University of Utah, Salt Lake City, UT 84105, USA
}

\date{Accepted 2023 September 19. Received 2023 September 12; in original form 2023 June 20}

\pubyear{2023}

\graphicspath{{./}{figures/}}
\begin{document}
\label{firstpage}
\pagerange{\pageref{firstpage}--\pageref{lastpage}}\maketitle

\begin{abstract}
Previous results in the literature have found the young inner-disk open cluster NGC 6705 to be mildly $\alpha$-enhanced. We examined this possibility via an independent chemical abundance analysis for 11 red-giant members of NGC 6705. The analysis is based on near-infrared APOGEE spectra and relies on LTE calculations using spherical model atmospheres and radiative transfer. We find a mean cluster metallicity of $\rm [Fe/H] = +0.13 \pm 0.04$, indicating that NGC 6705 is metal-rich, as may be expected for a young inner-disk cluster. The mean $\alpha$-element abundance relative to iron is $\rm \langle [\alpha/Fe]\rangle =-0.03 \pm 0.05$, which is not at odds with expectations from general Galactic abundance trends. 
NGC 6705 also provides important probes for studying stellar mixing, given its turn-off mass of M$\sim$3.3 M$_\odot$. Its red giants have low $^{12}$C abundances ([$^{12}$C/Fe]=$-$0.16) and enhanced $^{14}$N abundances ([$^{14}$N/Fe]=+0.51), which are key signatures of the first dredge-up on the red giant branch. An additional signature of dredge-up was found in the Na abundances, which are enhanced by [Na/Fe]=+0.29, with a very small non-LTE correction. The $^{16}$O and Al abundances are found to be near-solar. All of the derived mixing-sensitive abundances are in agreement with stellar models of approximately 3.3 M$_{\odot}$ evolving along the red giant branch and onto the red clump.
As found in young open clusters with similar metallicities, NGC 6705 exhibits a mild excess in the s-process element cerium, with $\rm [Ce/Fe] = +0.13\pm0.07$. 
\end{abstract}


\section{Introduction} \label{sec:intro}

Open clusters are excellent probes of chemical evolution in the Milky Way disk as their range in metallicity overlaps that of the disk field stars, while their locations extend from the inner disk to the outskirts of the Galaxy, and they have formed over an extended period of Galactic history.
NGC 6705 (Messier 11, M11) is relatively star-rich, compact, and located in the inner disk at a Galactocentric distance of 6.5 kpc \citep{cantat2020A&A...640A...1C}. This open cluster is young, with a well-defined age from isochrone fitting of $316 \pm 50$ Myr \citep{Cantat2014,dias2021MNRAS.504..356D} and, given its youth, has probably not migrated very far from its birthplace, with an estimated birth radius between 6.8 – 7.5 kpc \citep{Casamiquela2018AA...610A..66C}.

NGC 6705 has been reported to be a metal-rich cluster, with [Fe/H]$\sim$+0.1--0.17 \citep[e.g.;][]{Gonzalez2000,Heiter2014,Magrini2017,Casamiquela2018AA...610A..66C}, although recent works have also found it to have mean metallicities closer to solar \citep{magrini2021A&A...651A..84M,casamiquela2021A&A...652A..25C,randich2022A&A...666A.121R}. 
Another aspect of this open cluster is that previous studies in the literature \citep{Casamiquela2018AA...610A..66C,Tautvaiviense2015} have found its stellar members to be moderately enhanced in $\alpha$-elements relative to iron ([$\alpha$/Fe] $\sim +0.1$ to +0.2), although it has an age of only a few hundred million years. \cite{Magrini2014} also found NGC 6705 to be mildly enhanced in $\alpha$-elements: <[$\alpha$/Fe]> = +0.08, while the more recent study of \cite{Magrini2017} found that only [Mg/Fe], and not the other $\alpha$-elements, was enhanced (by +0.1 dex) in NGC 6705.

Youth and $\alpha$-enhancement together are not expected from simple chemical evolution modeling, as $\alpha$-elements, such as O, Mg, Si, and Ca, are produced at early times, mainly in Type II Supernovae (formed by massive stars on short timescales); enrichment in the [$\alpha$/Fe] ratio generally indicates that a star formed from gas enriched by SN II before SN Ia had the time to contribute iron to the natal gas.  Within the Galactic disk, there are two sequences defined by the $\alpha$-element abundances: the high- and the low-$\alpha$ sequences \citep[e.g.,][]{fuhrmann1998A&A...338..161F,Reddy2006,nidever2014ApJ...796...38N,anders2014AA...564A.115A,Hayes2018ApJ...852...49H,queiroz2020AA...638A..76Q};
the high-$\alpha$ sequence is older and corresponds to the thick disk (rich in $\alpha$-elements relative to its Fe content), while the thin disk population consists of stars with lower values of [$\alpha$/Fe].

There are, however, interesting results in the literature pointing to a population of young field stars with a high abundance ratio of $\alpha$-element-to-iron, which are unusual given their ages reported in the literature.
\cite{Chiappini2015} discovered such young [$\alpha$/Fe]-enhanced stars in a sample of field stars observed by the SDSS APOGEE Survey \citep{Majewski2017} having CoRoT asteroseismology \citep{baglin2006ESASP1306...33B}. CoRoT provided precise age estimation for field stars, resulting in the identification of a large number of young stars in the inner region of the Galactic disk that are rich in $\alpha$-elements ($[\alpha/Fe] \sim 0.1 - 0.3$) and with a low abundance of iron-peak elements. 
Meanwhile, \cite{Martig2015} analyzed a sample of 1639 red giants with astereoseismic ages from the APOGEE sample and observed by the Kepler mission (referred to as the APOKASC sample) to investigate the relationship between age and chemical abundances. As a result of their analysis, they identified fourteen stars enriched in $\alpha$-elements ([$\alpha$/Fe]$>$0.13) that were younger than 6 Gyr, and five stars with [$\alpha$/Fe]$\geq$ 0.20 are younger than 4 Gyr. Possible scenarios to explain this young, $\alpha$-enhanced population of stars include accretion of material from a binary companion or binary mergers \citep[e.g.,][]{izzard2018MNRAS.473.2984I,silvaaguirre2018MNRAS.475.5487S,hekker2019MNRAS.487.4343H,jofre2023AA...671A..21J}, resulting in stars that would appear to be young, while actually being old. In addition, \cite{miglio2021AA...645A..85M} found that stars with [$\alpha$/Fe] $>$ 0.1 from the Kepler field that appeared young were overmassive; this result supports the scenario that most of these stars have undergone an interaction with a companion.

Given its youth, coupled to the mild $\alpha$-enhancements found for NGC 6705 in the literature \citep{Casamiquela2018AA...610A..66C,Tautvaiviense2015}, along with the different signatures ($\alpha$-enhanced versus non $\alpha$-enhanced) obtained, for example, for Mg in comparison with other $\alpha$-elements \citep{Magrini2017}, and the relevance of finding a young $\alpha$-rich open cluster in the context of the young $\alpha$-enhanced field stars in the Galaxy \citep[as discussed in][]{Casamiquela2018AA...610A..66C}, it becomes important to revisit the $\alpha$-abundances in NGC 6705 from a completely independent analysis, also keeping in mind that the results in the literature for NGC 6705 mentioned above are all from optical studies. All, except for \cite{Casamiquela2018AA...610A..66C}, being based on the Gaia-ESO survey \citep[GES;][]{Gilmore2012,gilmore2022A&A...666A.120G,randich2022A&A...666A.121R}. 

In addition, due to its relative youth and stellar richness in comparison to other open clusters, NGC 6705 contains a populous sample of red giants in which to probe stellar mixing in the interesting mass range between M$\sim 3.0 - 3.5$ M$_{\odot}$.  Such intermediate-mass red giants can exhibit measurable chemical abundance changes due to deep mixing beyond the usual variations in $^{12}$C, $^{13}$C, and $^{14}$N observed in lower-mass red giants, to include possible changes in the $^{16}$O, or Na, or Al abundances.  The red giants in NGC 6705 can provide an important observational test of stellar models.
\cite{smiljanic2016AA...589A.115S} studied Na and Al in low- and intermediate-mass clump giants, in particular in six open clusters from the Gaia ESO survey, and found both Na and Al to be enriched in NGC 6705. While their Na results for this cluster were in agreement with predictions from stellar evolution models, their Al abundances were above model predictions, as aluminum is not expected to be affected by mixing in the mass range of NGC 6705 giants \citep{lagarde2012AA...543A.108L}.

In this study, we select a sample of red-giant stars which are members of NGC 6705 in order to determine their stellar parameters and present a detailed analysis of the chemical abundances of their $\alpha$-elements (O, Mg, Si, and Ca, and Ti), along with iron and the Fe-peak elements (V, Cr, Mn, Co, and Ni), elements sensitive to red giant mixing ($^{12}$C, $^{14}$N, Na, and Al), as well as the s-process element cerium. 
This spectroscopic analysis is based on APOGEE spectra, which are in the near-infrared, but uses an independent analysis and methodology when compared with the APOGEE abundance pipeline ASPCAP \citep{Garcia2016}, particularly in the derivation of the stellar parameters effective temperature and surface gravity, and given the well-known systematic offsets in surface gravity values for the ASPCAP results, which are post-calibrated \citep{jonsson2020AJ....160..120J}. 
This paper is organized as follows: in Section \ref{sec:sample}, we present the sample and the observations, while Section \ref{sec:analysis} describes the methodology to determine the stellar parameters, and Section \ref{sec:indiv_abun} presents the individual abundance analysis of seventeen elements. In Section \ref{sec:compar}, we compared our results with literature results. Section \ref{sec:res_disc} contains a discussion of the results, and Section \ref{sec:conclusions} the conclusions.

\section{Observation and Sample} \label{sec:sample}

\subsection{APOGEE Spectra}
The Apache Point Observatory Galactic Evolution Experiment \citep[APOGEE;][]{Majewski2017} was one of the three surveys carried out as part of the Sloan Digital Sky Survey-IV \citep[SDSS-IV;][]{Blanton2017}. APOGEE targeted the open cluster NGC 6705 as part of its OCCAM (Open Cluster Chemical Abundances and Mapping) campaign, which aimed to study the structure and chemical evolution of the Milky Way \citep{Frinchaboy2013,donor2018,myers2022AJ....164...85M}. 
The APOGEE spectra analyzed in this study were obtained using a 300-fiber cryogenic spectrograph on the 2.5 m telescope at the Apache Point Observatory \citep[New Mexico, USA;][]{Gunn2006} and these have a resolution R $ = \lambda/\Delta\lambda \sim$ 22,500 and spectral coverage from 1.51 to 1.69 $\mu$m \citep{Wilson2010,wilson2019PASP..131e5001W}. Reduction of the APOGEE spectra, as well as the determination of the stellar radial velocities, were carried out by an automated data processing pipeline \citep{Niveder2015}, and the reduced spectra analyzed here come from the publicly available 17th APOGEE data release \citep[DR17;][]{abdurrouf2022ApJS..259...35A}. 
The open cluster NGC 6705 was observed in APOGEE field 027-04, identified by location ID 4470 \citep{Zasowski2013,zasowski2017AJ....154..198Z,beaton2021AJ....162..302B}. APOGEE targeted a total of 343 stars in this field which were investigated here for cluster membership in the section below. 

\subsection{NGC 6705 Membership}
The open cluster NGC 6705 is located at Galactic coordinates $l = 27.304^o$ and $b=-2.773^o$, at an estimated distance of $\sim$1900 -- 2200 pc \citep[e.g.,][]{CantatGaudin2020A&A...633A..99C,dias2021MNRAS.504..356D,hunt2023arXiv230313424H}. The mean radial velocity of its cluster members, according to radial velocities from Gaia DR2, was estimated in  \citet{dias2021MNRAS.504..356D} to be $35.68 \pm 0.24$ \kms and considering 357 stars \cite{tarricq2021A&A...647A..19T} found a mean radial velocity of $34.49 \pm 0.27$ \kms. 

\subsubsection{Membership according to HDBSCAN} \label{sec:memb}
We used the python code HDBSCAN \citep[Hierarchical Density-Based Spatial Clustering of Applications with Noise;][]{campello10.1007/978-3-642-37456-2_14} 
clustering algorithm to independently assess which stars from the observed APOGEE field 027-04 would be identified as members of the NGC 6705 open cluster. HDBSCAN is an unsupervised machine learning method, which does not require learning from labeled data to make predictions. More specifically, HDBSCAN is a density-based clustering algorithm that groups data points together based on their proximity and density.

HDBSCAN uses a number of input parameters that can be adjusted to control its clustering estimations. The main input parameters are: \textit{min\_cluster\_size}, \textit{min\_samples}, \textit{cluster\_selection\_epsilon} and \textit{alpha}.  The parameter \textit{min\_cluster\_size} sets the minimum number of points needed to define a distinct cluster, thus any potential cluster that might contain fewer points would be labeled as noise (outliers). 
The \textit{min\_samples} parameter defines the minimum number of neighboring points surrounding a given point for it to be considered as a core point, while a minimum distance below which HDBSCAN will not further split a cluster is set by the parameter \textit{cluster\_selection\_epsilon}. 
The parameter \textit{alpha} controls the balance between condensed tree density and hierarchy depth. Higher values of \textit{alpha} result in clusters that are more tightly bound within the hierarchy, while lower values allow clusters to be more easily split. For the APOGEE field 027-04, we used the following parameter values: \textit{min\_cluster\_size$=$4}, \textit{min\_samples$=$3}, \textit{cluster\_selection\_epsilon$=$0.0}, and \textit{alpha$=$1.0}.

\begin{figure}
\centering
\includegraphics[width=0.52\textwidth]{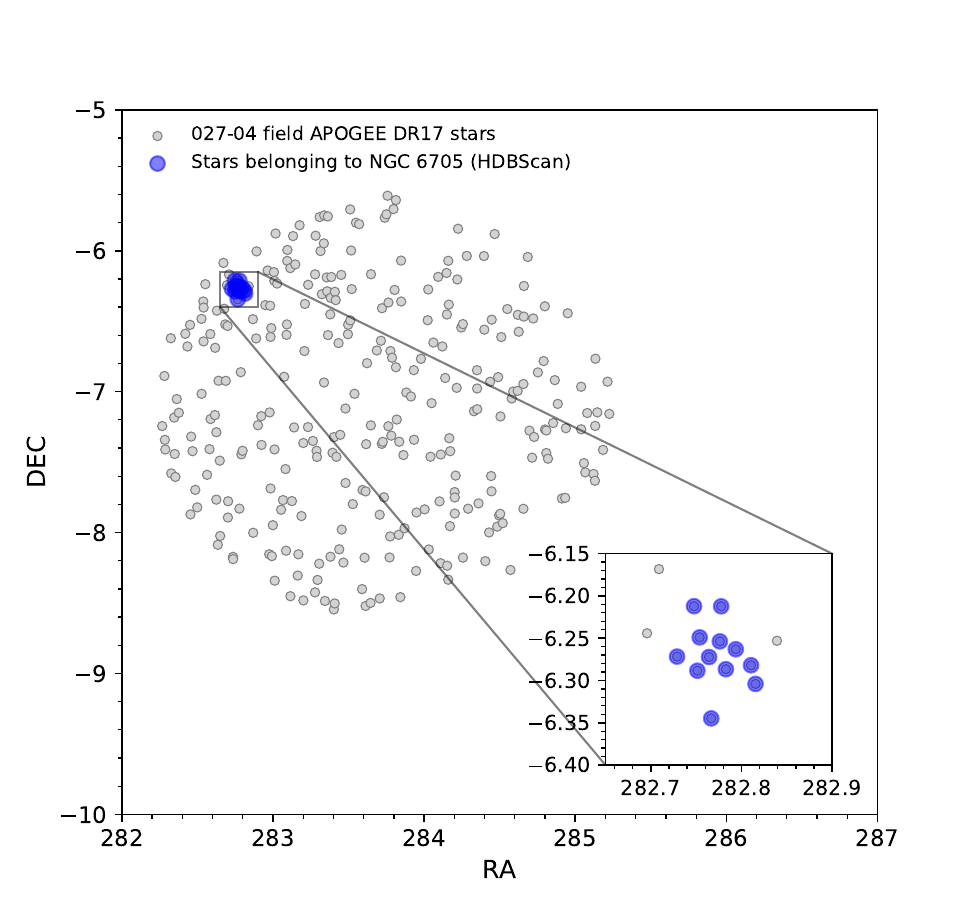}
\caption{Declination (DEC) versus right ascension (RA) of stars from 027-04 field - APOGEE DR17. The stars highlighted in blue are the stars pointed out as members of the NGC 6705 cluster using the HDBSCAN code. The highlighted panel is a region zoom containing these stars.}
\label{fig:ra_dec}
\end{figure}

To search for cluster members within the observed stars in the APOGEE plate, we used three parameters: proper motions (pm RA and pm DEC) from Gaia DR3 \citep[Gaia DR3;][]{gaia2021A&A...649A...1G}, distances from \cite{bailer-jones2021} and radial velocities from APOGEE DR17. 
As mentioned above, the APOGEE 027-04 field has 343 stars, however, not all of them have all three parameters available.
Figure \ref{fig:ra_dec} shows the RA-DEC space of the 307 stars analyzed with the HDBSCAN code. From this group, twelve stars have been selected as cluster members according to our parameters and these are shown as blue circles in Figure \ref{fig:ra_dec}. 
The results from HDBSCAN indicated that eleven of the stars had 100 \% probability of being cluster members, while the star 2MASS J18510399-0620414 had a slightly lower probability (93 \%) of belonging to the cluster. 

\subsubsection{Other Membership from the Literature}

\begin{figure*}
\centering
\includegraphics[width=0.32\textwidth]{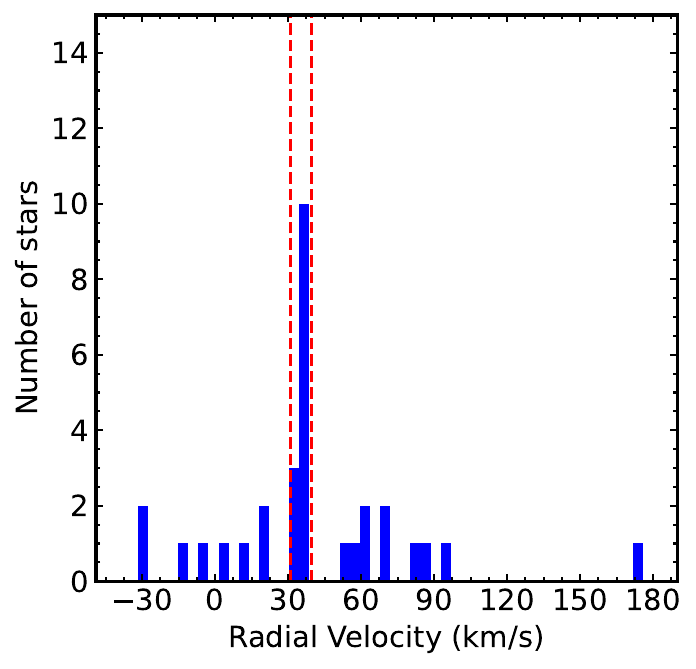}
\includegraphics[width=0.32\textwidth]{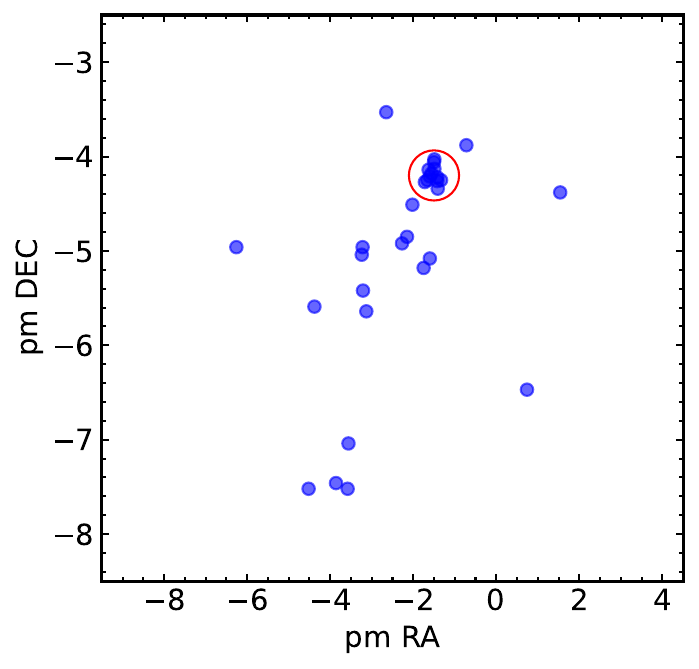}\\
\includegraphics[width=0.325\textwidth]{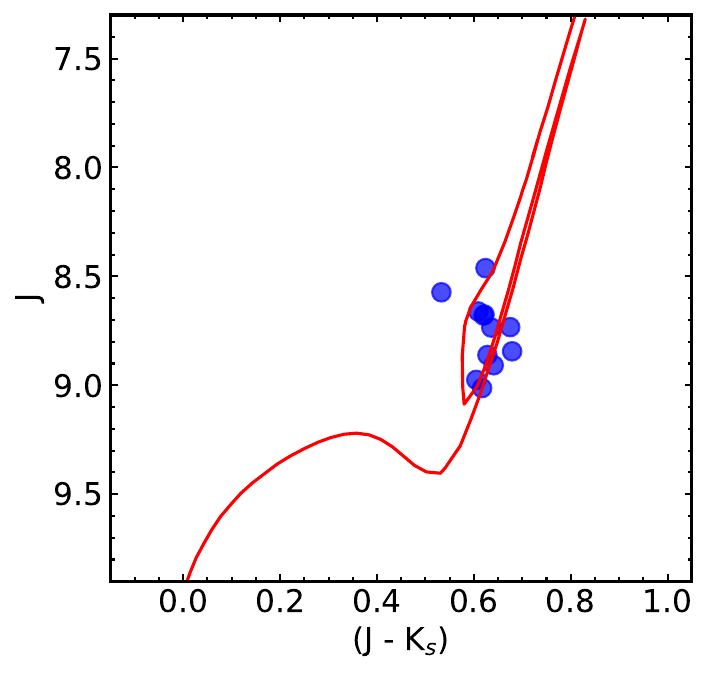}
\includegraphics[width=0.32\textwidth]{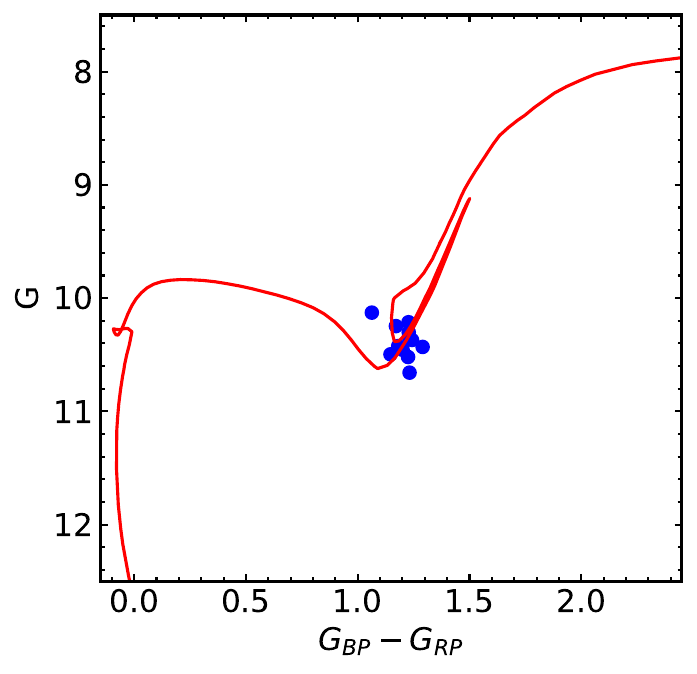}
\caption{The upper left panel displays the radial velocity distribution of stars in the 027-04 field, sourced from APOGEE DR17. The red dashed lines correspond to the radial velocity range determined for the cluster. In the upper right panel, the Gaia DR3 proper motions of the targeted stars are presented. The blue dots inside the red circle signify stars within the radial velocity range determined for the cluster. The bottom panels depict the 2MASS ($J-K_s$) vs. $J$ and Gaia DR2 ($G_{BP} - G_{RP}$) vs. $G$ diagrams. The red line in both cases represents the isochrones from Bressan et al. (\citeyear{Bressan2012}).}
\label{fig:membership6705}
\end{figure*}

We also investigated which stars observed by APOGEE would be members of NGC 6705 according to other studies in the literature.
Thirty one stars have been labeled as possible members of NGC 6705 in the most recent OCCAM study by \citep{myers2022AJ....164...85M} (their Table 3) and the upper right panel of Figure \ref{fig:membership6705} shows the proper motions of these stars from Gaia DR3 \citep{gaia2021A&A...649A...1G}. Their APOGEE radial velocities are shown in the left panel of Figure \ref{fig:membership6705} and it can be seen that these show a large variation in radial velocities, with RV values ranging mostly between $-$30--90 km/s. 
From this sample of 31 stars, \cite{myers2022AJ....164...85M} used APOGEE radial velocities and metallicities, along with Gaia proper motions to estimate membership probabilities to select a sample of twelve stars that were considered to be members of NGC 6705. In Figure \ref{fig:membership6705} the twelve member stars are depicted inside the red circle in right panel and within the red dashed lines in left panel.

We also verified which stars observed by APOGEE would be members of NGC 6705 according to the probabilities of membership provided by \cite{CantatGaudin2020A&A...633A..99C}. There are 1183 stars in that catalog with a probability of cluster membership larger or equal to 0.7. The cross-match of their member list with the APOGEE DR17 database led to the identification of twelve stars in common.  In addition, these same 12 stars are considered as members by \cite{dias2021MNRAS.504..356D} and eleven of them are members according to \cite{jackson2022MNRAS.509.1664J}.

In summary, we independently identified a sample of twelve bona fide stellar members of NGC 6705 and this membership is in agreement with the results from other independent studies in the literature. The NGC 6705 members are presented in Table \ref{tab:sample}, along with the star's 2MASS J, H, K$_s$ \citep[Two Micron All Sky Survey;][]{Cutri2003}, and V magnitudes taken from \cite{Cantat2014}, \cite{Tautvaiviense2015}, \cite{casamiquela2016MNRAS.458.3150C} and \cite{Zacharias2005}
and Gaia magnitudes (G, G$_{BP}$ and G$_{RP}$) from Gaia DR3 \citep[][]{gaia2021A&A...649A...1G}.

Also included are the radial velocities (RV), along with the RV dispersions derived from individual APOGEE visits, and the signal-to-noise of the APOGEE spectra. We note that the star 2MASS J18510092-0614564 is deemed to be a binary given the scatter in the Gaia radial velocity measurements (Gaia DR3 RV=37.45 $\pm$ 11.37 km/s) and will not be analyzed in this study.

Finally, in the two lower panels of Figure \ref{fig:membership6705} we show the ($J-K_s$) vs. $J$ and ($G_{BP} - G_{RP}$) vs. $G$ diagrams using 2MASS and Gaia DR3 photometry for the sample of twelve stars. The red line in both panels represents the PARSEC isochrone \citep{Bressan2012} for the age and metallicity of NGC 6705 (0.316 Gyr and 0.10 dex, respectively). The location of the stars relative to the isochrones in the color magnitude diagrams presented indicates that the selected members are probably red-clump stars, although they could also be on the red-giant branch.

\begin{table*}
\centering
\caption{Member stars of NGC 6705}
\begin{tabular}{lcccccccccc}
\hline \hline
Star ID  & J & H & K & V & G & G$_{BP}$ & G$_{RP}$ & RV & SNR \\
& (mag) & (mag) & (mag) & (mag) & (mag) & (mag) & (mag) & (km/s) &  \\ \hline
2M18505494-0616182 & 9.199 & 8.498 & 8.318 & 11.860 & 11.338 & 12.131 & 10.467 & 35.019 $\pm$ 0.059 & 400 \\
2M18510399-0620414 & 9.090 & 8.400 & 8.252 & 11.872 & 11.313 & 12.164 & 10.411 & 34.623 $\pm$ 0.018 & 467 \\
2M18510661-0612442 & 9.035 & 8.406 & 8.213 & 11.720 & 11.184 & 11.997 & 10.306 & 33.761 $\pm$ 0.047 & 462 \\
2M18511048-0615470 & 8.817 & 8.224 & 7.991 & 11.627 & 11.095 & 11.905 & 10.214 & 33.597 $\pm$ 0.019 & 488 \\
2M18505944-0612435 & 9.330 & 8.722 & 8.523 & 11.872 & 11.378 & 12.138 & 10.528 & 34.710 $\pm$ 0.022 & 385 \\
2M18510032-0617183 & 9.368 & 8.751 & 8.549 & 12.081 & 11.540 & 12.347 & 10.652 & 36.689 $\pm$ 0.050 & 396 \\
2M18510341-0616202 & 9.216 & 8.579 & 8.386 & 11.801 & 11.305 & 12.087 & 10.443 & 37.108 $\pm$ 0.007 & 429 \\
2M18510786-0617119 & 9.030 & 8.399 & 8.206 & 11.621 & 11.129 & 11.905 & 10.271 & 34.572 $\pm$ 0.019 & 359 \\
2M18511452-0616551 & 9.263 & 8.620 & 8.420 & 11.923 & 11.402 & 12.209 & 10.521 & 36.307 $\pm$ 0.045 & 306 \\
2M18510092-0614564 & 9.016 & 8.395 & 8.205 & 11.484 & 11.010 & 11.716 & 10.189 & 34.827 $\pm$ 0.509 & 466 \\
2M18510626-0615134 & 8.928 & 8.379 & 8.193 & 11.627 & 11.222 & 12.030 & 10.337 & 35.116 $\pm$ 0.007 & 449 \\
2M18511571-0618146 & 9.088 & 8.445 & 8.211 & 11.807 & 11.252 & 12.073 & 10.367 & 35.442 $\pm$ 0.024 & 475 \\
\hline
\end{tabular}
\label{tab:sample}
\end{table*}


\section{Stellar Parameters} \label{sec:analysis}
The determination of the abundances of chemical elements from stellar spectra relies on fundamental stellar atmospheric parameters, such as the effective temperature ($T_{\rm eff}$), surface gravity (log $g$), microturbulence velocity ($\xi$), and metallicity ([Fe/H]). To derive these stellar parameters, we employed a methodology that is similar to the analysis presented in \cite{Souto2016}.


Stellar effective temperatures were derived from the 2MASS \citep[Two Micron All Sky Survey;][]{Cutri2003} magnitudes J, H and K$_s$, and the photometric calibrations of \cite{Gonzalez2009} through the equation:
\begin{equation}\label{eq:Teff}  
    \theta _{eff} = b_0 + b_1 X + b_2 X^2 + b_3 X[Fe/H] + b_4 [Fe/H] + b_5 [Fe/H]^2 
\end{equation} where T$\rm _{eff} = 5040/\theta _{eff}$, the values $X$ represent the colors $V-J$, $V-H$, $V-K_s$, and $J-K_s$, and the constants $b_0$, $b_1$, $b_2$, $b_3$, $b_4$, and $b_5$ for each photometric color can be found in \citet{Gonzalez2009}. The adopted metallicity in this step was [Fe/H]=0.10 dex (taken from \cite{Cantat2014}).

The reddening value adopted in this study was $E(B-V)=0.4$ \citep{Cantat2014} and reddening corrections were computed using the relations in \cite{Bilir2008}. 
Table \ref{tab:params}  lists the effective temperatures determined from each color and the corresponding median effective temperatures (and median absolute deviation, MAD). The effective temperatures obtained from different colors agree quite well, with the MAD for most stars being less than 50 K, which is a typical uncertainty for effective temperature scales. We note also that these errors are similar to those found in \cite{Souto2016} for a sample of red-giants in the open cluster NGC 2420.

\begin{table*}
\centering
\caption{Atmospheric Parameters}
\begin{tabular}{lcccccccc}
\hline \hline
Star ID & $T_{\rm eff}$(V-J) & $T_{\rm eff}$(V-H) & $T_{\rm eff}$(V-K$_s$) & $T_{\rm eff}$(J-K$_s$) & $ \langle T_{\rm eff}\rangle$ & log $g$ & [Fe/H] & $\xi$ \\
 & (K) & (K) & (K) & (K) & (K) & (cm/s$^2$) & (dex) & (km/s)  \\ \hline
2M18505494-0616182 & 4735 & 4692 & 4715 & 4632 & 4704 $\pm$ 31 
& 2.279 & 0.15 & 1.50 \\
2M18510399-0620414 & 4599 & 4601 & 4651 & 4754 & 4626 $\pm$ 51 
& 2.253 & 0.07 & 1.70 \\
2M18510661-0612442 & 4707 & 4734 & 4745 & 4801 & 4739 $\pm$ 27 
& 2.250 & 0.14 & 1.80 \\
2M18511048-0615470 & 4570 & 4658 & 4638 & 4789 & 4648 $\pm$ 63 
& 2.167 & 0.12 & 1.60 \\
2M18505944-0612435 & 4882 & 4884 & 4885 & 4846 & 4883 $\pm$ 14 
& 2.386 & 0.16 & 1.60 \\
2M18510032-0617183 & 4675 & 4720 & 4723 & 4810 & 4722 $\pm$ 39 
& 2.359 & 0.03 & 1.70 \\
2M18510341-0616202 & 4827 & 4816 & 4825 & 4778 & 4821 $\pm$ 17 
& 2.334 & 0.15 & 1.75 \\
2M18510786-0617119 & 4819 & 4816 & 4825 & 4795 & 4818 $\pm$ 9 
& 2.272 & 0.17 & 1.90 \\
2M18511452-0616551 & 4736 & 4744 & 4748 & 4740 & 4742 $\pm$ 4 
& 2.318 & 0.15 & 1.70 \\
2M18510626-0615134 & 4691 & 4793 & 4808 & 5073 & 4800 $\pm$ 116 
& 2.304 & 0.16 & 1.80 \\
2M18511571-0618146 & 4668 & 4692 & 4670 & 4643 & 4669 $\pm$ 13 
& 2.263 & 0.18 & 1.50 \\
\hline
\end{tabular}
\label{tab:params}
\end{table*}

\begin{figure}
\centering
\includegraphics[width=0.4\textwidth]{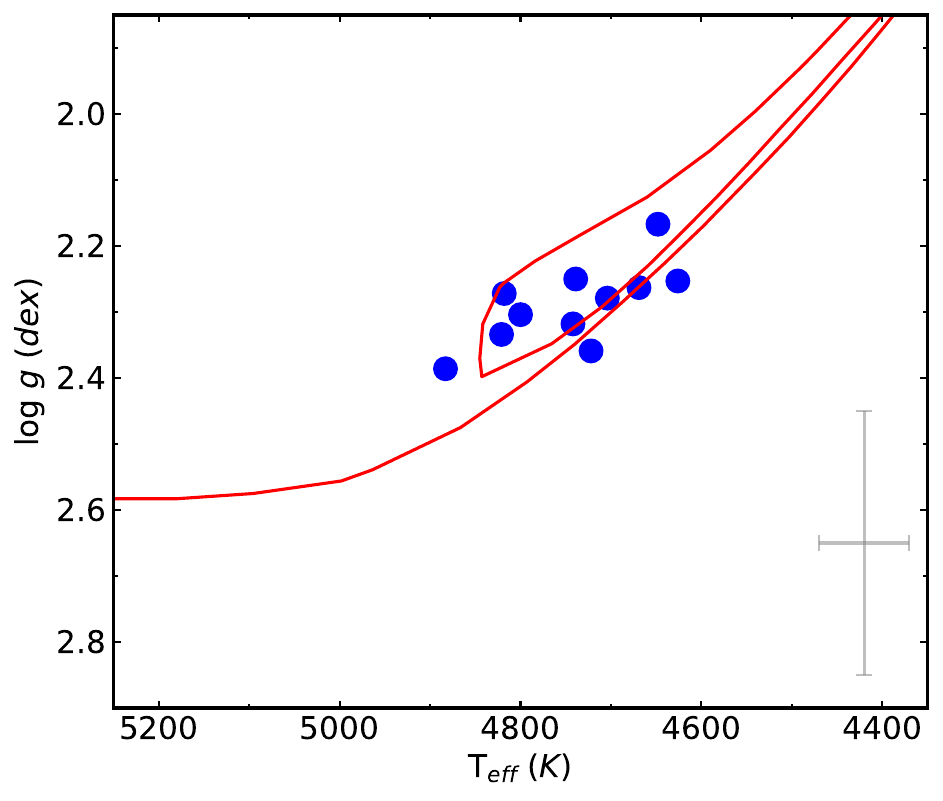}
\caption{Kiel diagram. The red line in the plot represents the isochrone for metallicity of $\rm [Fe/H] = 0.13$ dex and age of 0.316 Gyr. The isochrone was calculated using PARSEC \citep{Bressan2012}. Effective temperatures and surface gravities are the values determined in this work (see Table \ref{tab:params}). A typical errorbar is shown in the bottom right of the figure.}
\label{fig:path_evol}
\end{figure}

To determine the surface gravities for the targets, the fundamental relation (equation \ref{eq:log}) was used, with the following reference solar parameter values: log $g = 4.438$ (cgs), $T_{\rm eff,\odot}=5770$K, and a bolometric magnitude of $M_{bol,\odot}=4.75$ \citep{prsa2016}.  The effective temperatures used are the median effective temperatures listed in Table \ref{tab:params}. Stellar masses were derived using the PARSEC isochrones \citep{Bressan2012}, which yield a mass of $\sim$3.3 M$_\odot$ for a cluster age of 0.316 Gyr \citep{ Cantat2018cat} and a metallicity of $[M/H]=+0.10$ dex. Absolute magnitudes were determined using the distance module (m-M)$_o$ = 11.38 \citep{dias2021MNRAS.504..356D}, along with bolometric corrections from \cite{Montegriffo1998}:
\begin{equation}\label{eq:log}
 log g =log g_\odot +log \left(\frac{M_\star}{M_\odot}\right)+4log \left(\frac{T_\star}{T_\odot}\right)+0.4(M_{bol,\star}- M_{bol,\odot}).
\end{equation} 
The stellar parameters of our sample are presented in Figure \ref{fig:path_evol} as a Kiel diagram, where we also show as a red line a PARSEC isochrone \citep{Bressan2012} computed for a metallicity of 0.13 dex and an age of 0.316 Gyr. 
The studied stars occupy a small range in parameter space close to the red clump, with the effective temperatures ranging between $\sim$ 4600 -- 4900 K and surface gravities spanning from 2.2 to 2.4 dex. In Figure \ref{fig:path_evol} the stellar parameters seem to segregate, with four targets being hotter than 4800 K and more in line with them being on the red clump and seven targets cooler than 4800 K and falling closer to the RGB branch. However, given the uncertainties both in the stellar parameters and in the models, a secure distinction between red clump and RGB is difficult to make since we do not have information from asteroseismology.

\section{Abundance Analyses and Methodology} \label{sec:indiv_abun}

The chemical abundances for seventeen elements were calculated by comparing observed and synthetic spectra through the $\chi ^2$-fitting method. Synthetic spectra were generated using  MARCS model atmospheres \citep{gustafson2008A&A...486..951G} and the Brussels Automatic Code for Characterizing High Accuracy Spectra \citep[BACCHUS;][]{masseron2016ascl.soft05004M}, which utilizes the radiative transfer code Turbospectrum \citep{alvarez1998A&A...330.1109A,plez2012ascl.soft05004P}. The APOGEE line list was adopted in the calculations of synthetic spectra \citep{smith2021AJ....161..254S}.

We derived the stellar metallicities and microturbulent velocities using nine Fe I lines selected in the APOGEE region by \cite{Smith2013} and the methodology discussed in \cite{Souto2016}. 
Briefly, the methodology consists of measuring the iron abundance of each Fe I line for different values of microturbulent velocities ($\xi$) using the spectrum synthesis method. The adopted values of microturbulent velocities were the ones that produced the smallest spread between the iron abundances of the individual lines. 

The APOGEE spectra of red-giant stars are characterized by the presence of numerous molecular features, predominantly spectral lines from CO, CN, and OH, making it an ideal tool for accurately determining the abundances of carbon, nitrogen, and oxygen. The molecular lines used for abundance determination are $^{12}$C$^{16}$O, $^{12}$C$^{14}$N, and $^{16}$OH, respectively, with the procedure to determine the C, N, and O abundances following the \cite{Smith2013} methodology. First, we derive the abundance of carbon from the molecular CO lines, then the oxygen abundance from the OH lines, and lastly, the nitrogen abundance from the CN lines.

The spectral range covered by APOGEE also contains atomic lines from many elements, including $\alpha$-elements, such as Mg, Si, Ca, and Ti, along with odd-Z elements, such as, Na, Al and K, as well as the Fe-peak elements V, Cr, Mn, Fe, Co, and Ni, and the s-process element Ce. In this analysis, we analyzed 73 spectral features that had been selected in the previous APOGEE studies of red giants of \cite{Smith2013}, \cite{Cunha2015}, and \cite{Souto2016} as abundance indicators: 9 Fe I lines, 4 CO lines, 9 CN lines, 2 Na I lines, 6 Mg I lines, 2 Al I lines, 7 Si I lines, 2 K I lines, 3 Ca I lines, 5 Ti I lines, 1 V I line, 1 Cr I line, 3 Mn lines, 1 Co I line, 7 Ni I lines, 7 Ce II lines, and 4 OH lines, noting that the latter are blended with CN at the studied range in parameter space. 
Table \ref{tab:abundances} contains the atomic and molecular line list used in the abundance analysis, the corresponding line-by-line abundances, the mean abundances, and standard deviations for each star, while in Table \ref{tab:mean_abu} we list the mean abundances, and standard deviation ($STD$) obtained for the cluster, along with mean [X/H] ratios relative to the Solar abundances \citep{asplund2021A&A...653A.141A}.

\subsection{Abundance Sensitivities and Uncertainties} \label{sec:sensitive}

\citet[][their Table 4]{Souto2016} estimated abundance uncertainties due to changes in stellar parameters), $\sigma$, for all elements analyzed here, except for Ce. The uncertainties were computed by using the quadrature sum of abundance changes obtained by varying, respectively, the effective temperature by +50 K, the surface gravity by +0.2 dex, the metallicity by +0.2 dex, and the microturbulent velocity by +0.2 km/s \citep[see also the discussion in,][]{Smith2013,Souto2016}. Here, we add the uncertainties calculated for cerium: the change in the Ce abundance due to a +50 K variation in $T_{\rm eff}$ is +0.06 dex, while for $\Delta$log g of +0.2 it is +0.06 dex, for $\Delta\xi$ of +0.2 km/s it is -0.04 dex, and for $\Delta$[M/H] of +0.2 dex is +0.04 dex. Summing these abundance changes in quadrature we obtain an uncertainty in the Ce abundance of 0.1 dex. Table \ref{tab:mean_abu} (last column) lists these estimated uncertainties for all elements.

The elemental abundances in this study were derived, in general (except for V, Cr, and Co), from more than one atomic or molecular line, with the spread in the individual line abundances for a given star used to evaluate the internal consistency between the different line measurements. The standard deviations of the mean abundances in Table \ref{tab:abundances} are, for some elements quite small, being less than 0.04 -- 0.05 dex, such as for C and Ca, while for many of the elements it is $\sim$0.09 -- 0.1 dex, such as for O, N, Al, Fe, and Mg, while, in a few cases, the standard deviations of the mean reach values of  0.12 -- 0.15 dex in some stars, such as for Si, Ti, and Ce. 

Finally, given that the members of NGC 6705 presumably formed as a single stellar population, and that they are not affected by diffusion effects as they are on RGB or red clump \citep{souto2018ApJ...857...14S, bertelli2018MNRAS.478..425B,xudong2018MNRAS.481.2666G}, one can use the standard deviations of the mean elemental abundances in the cluster (Table \ref{tab:mean_abu}) to also gauge internal uncertainties in the analysis. The elements with the smallest abundance scatter among the member stars analyzed (STD$\leq 0.05$ dex) are O, N, Na, Mg, Fe, and Ni. 
The elements C, Al, Si, K, Ca, V, Mn, Co, and Ce exhibit higher scatter, although still moderate, ranging from 0.06 to 0.08 dex. For Cr the scatter is 0.09 dex (which can be related to the fact that this element has only one measurable weak line in the APOGEE window), while for Ti we find a scatter of 0.10 dex. 
We note that the standard deviation for the iron abundances (0.04 dex), for example, is comparable to that reported by \cite{Cunha2015} for the open cluster NGC 6791, and \cite{Souto2016} for NGC 2420, both of which used APOGEE spectra of red giants in their analyses.

\subsection{Non-LTE Corrections for Na, Mg, K, and Ca} \label{sec:non_etl}
Non-LTE corrections to the LTE Na, Mg, K, and Ca  abundances derived in this study can be estimated from the LTE and non-LTE abundances taken from spectral libraries generated for APOGEE DR17. Such synthetic spectra were computed using the Synspec spectral synthesis code \citep{hubeny2021arXiv210402829H}, the APOGEE line list \citep{smith2021AJ....161..254S}, APOGEE MARCS models \citep{gustafson2008A&A...486..951G}, and in the case of non-LTE, adopting atomic models for Na, Mg, K, and Ca discussed in \cite{osorio2020AA...637A..80O}. Both the LTE and non-LTE abundances were calculated using the ASPCAP pipeline \citep{Garcia2016}. 
The mean differences between the non-LTE and LTE abundance results for our sample stars are given in Table \ref{tab:non_lte}.

The mean abundance differences ``non-LTE -- LTE'' are quite small for all four elements. For sodium, magnesium, and calcium the corrections were found to be negative, indicating that the LTE abundances of these elements are slightly overestimated, but not significantly so, relative to the non-LTE abundances by $-0.02$, $-0.03$, and $-0.03$, respectively. The mean difference for potassium is positive but also insignificant, at +0.01. It's worth noting that these differences are within the uncertainties associated with the abundances of these elements (see Table \ref{tab:mean_abu}) and non-LTE corrected abundances will not be considered in this study.

\begin{table}
\centering
\caption{Mean abundance differences of Na, Mg, K, and Ca (with STD) in Non-LTE and LTE.}
\begin{tabular}{lccc}
\hline \hline
Element & $\langle \delta [X/H]\rangle$ \\ 
        & (non-LTE - LTE) \\
\hline
Na      & $-0.019 \pm 0.003$ \\
Mg      & $-0.027 \pm 0.006$ \\
K       & $+0.007 \pm 0.010$ \\
Ca      & $-0.032 \pm 0.010$ \\
\hline
\end{tabular}
\label{tab:non_lte}
\end{table}
\clearpage
\onecolumn
{\small
\begin{landscape}
\renewcommand{\arraystretch}{0.95}
\begin{longtable}{llccccccccccc}
\caption{Line-by-line Elemental Abundances.} \label{tab:abundances} \\
\hline \hline
Element & $\lambda$ (\AA) & J18505494 & J18510399 & J18510661 & J18511048 & J18505944 & J18510032 & J18510341 & J18510786 & J18511452 &  J18510626 & J18511571 \\
 &  & -616182 & -620414 & -612442 & -615470 & -612435 & -617183 & -616202 & -617119 & -616551 & -615134 & -618146 \\ \hline
\endfirsthead
\multicolumn{13}{c}%
{{\bfseries \tablename\ \thetable{} -- continued}} \\
\hline \hline
Element & $\lambda$ (\AA) & J18505494 & J18510399 & J18510661 & J18511048 & J18505944 & J18510032 & J18510341 & J18510786 & J18511452 &  J18510626 & J18511571 \\
 &  & -616182 & -620414 & -612442 & -615470 & -612435 & -617183 & -616202 & -617119 & -616551 & -615134 & -618146 \\ \hline
\endhead
\hline \\
\endfoot
\hline 
\endlastfoot
C from CO  & 15580. & 8.42  &  8.33  &  8.41  &  8.45  &  8.51  &  8.33  &  8.49  &  8.51  &  8.43  &  8.45  &  8.43  \\
  &  15977. & 8.48  &  8.29  &  8.46  &  8.33  &  8.45  &  8.24  &  8.44  &  8.55  &  8.39  &  8.40  &  8.44  \\
  &  16186. & 8.47  &  8.35  &  8.55  &  8.42  &  8.54  &  8.32  &  8.51  &  8.60  &  ...   &  8.44  &  8.48  \\
  &  16613. & 8.51  &  8.33  &  8.47  &  8.39  &  8.50  &  8.31  &  8.50  &  8.51  &  8.45  &  8.49  &  8.48  \\
$\langle A(C) \rangle$ &  & $8.47 \pm 0.03$ & $8.32 \pm 0.02$ & $8.47 \pm 0.05$ & $8.40 \pm 0.04$ & $8.50 \pm 0.03$ & $8.30 \pm 0.04$ & $8.48 \pm 0.03$ & $8.54 \pm 0.04$ & $8.42 \pm 0.02$ & $8.45 \pm 0.03$ & $8.46 \pm 0.02$  \\
N from CN  & 15260. & 8.43 & 8.59 & 8.50 & 8.52 & 8.46 & 8.49 & 8.49 & 8.46 & 8.43 & 8.52 & 8.40  \\
  &  15322. & 8.47 & 8.60 & 8.49 & 8.56 & 8.50 & 8.46 & 8.53 & 8.53 & 8.49 & 8.53 & 8.39  \\
  &  15397. & 8.60 & 8.63 & 8.59 & 8.51 & 8.58 & 8.50 & 8.59 & 8.69 & 8.50 & 8.51 & 8.42  \\
  &  15332. & 8.50 & 8.54 & 8.49 & 8.56 & 8.55 & 8.57 & 8.55 & 8.54 & 8.49 & 8.53 & 8.45  \\
  &  15410. & 8.41 & 8.58 & 8.42 & 8.42 & 8.46 & 8.46 & 8.46 & 8.49 & 8.45 & 8.52 & 8.37  \\
  &  15447. & 8.42 & 8.63 & 8.48 & 8.55 & 8.53 & 8.51 & 8.52 & 8.55 & 8.56 & 8.52 & 8.48  \\
  &  15466. & ... &  ... &  8.28 & 8.40 & 8.42 & ... &  8.41 & 8.42 & 8.37 & 8.41 & 8.27  \\
  &  15472. & 8.29 & 8.47 & 8.33 & 8.35 & 8.34 & 8.40 & ... &  ... &  8.26 & 8.43 & 8.30  \\
  &  15482. & 8.36 & 8.52 & 8.38 & 8.40 & 8.45 & 8.39 & 8.48 & 8.46 & 8.33 & 8.50 & 8.33  \\
$\langle A(N) \rangle$ &  & $8.43 \pm 0.09$ & $8.57 \pm 0.05$ & $8.44 \pm 0.09$ & $8.47 \pm 0.08$ & $8.48 \pm 0.07$ & $8.47 \pm 0.05$ & $8.50 \pm 0.05$ & $8.52 \pm 0.08$ & $8.43 \pm 0.09$ & $8.50 \pm 0.04$ & $8.38 \pm 0.07$  \\
O from OH & 15280. & ... &  8.77 & 8.78 & 8.73 & 8.81 & ... &  8.86 & ... &  ... &  ...   8.83  \\
  &  15505. & 8.80 & 8.85 & 8.57 & 8.65 & 8.65 & 8.77 & 8.70 & 8.76 & 8.72 & 8.69 & 8.78  \\
  &  15570. & ... &  ... &  ... &  8.87 & ... &  ... &  ... &  ... &  ... &  ... &  ...  \\
  &  16191. & ... &  8.71 & 8.75 & 8.68 & 8.77 & ... &  8.81 & 8.85 & ... &  8.85 & 8.69  \\
$\langle A(O) \rangle$ &  & $8.80$ & $8.78 \pm 0.06$ & $8.70 \pm 0.09$ & $8.73 \pm 0.08$ & $8.75 \pm 0.07$ & $8.77$ & $8.79 \pm 0.07$ & $8.80 \pm 0.05$ & $8.72$ & $8.77 \pm 0.08$ & $8.77 \pm 0.06$  \\
Fe I &  15194.492 & 7.72 & 7.50 & 7.69 & 7.61 & ... &  7.50 & 7.72 & ... &  7.67 & 7.75 & 7.64  \\
  &  15207.526 & 7.58 & ... &  7.43 & 7.40 & ... &  ... &  ... &  ... &  7.47 & 7.60 & ...   \\
  &  15395.718 & 7.64 & 7.51 & 7.62 & 7.64 & 7.59 & 7.52 & 7.49 & 7.58 & 7.66 & 7.64 & 7.71  \\
  &  15490.339 & 7.65 & 7.53 & 7.61 & 7.53 & 7.50 & 7.42 & 7.59 & 7.54 & 7.53 & 7.58 & 7.61  \\
  &  15648.51 &  7.64 & 7.69 & 7.65 & 7.63 & 7.73 & 7.56 & 7.69 & 7.59 & 7.69 & 7.64 & 7.65  \\
  &  15964.867 & 7.50 & 7.42 & 7.62 & 7.58 & 7.56 & 7.48 & 7.63 & 7.65 & 7.60 & 7.60 & 7.65  \\
  &  16040.657 & 7.54 & 7.46 & 7.54 & 7.56 & 7.58 & 7.46 & 7.48 & 7.62 & 7.52 & 7.55 & 7.56  \\
  &  16153.247 & 7.53 & 7.49 & 7.52 & 7.60 & 7.61 & 7.45 & 7.54 & 7.68 & ... &  7.55 & 7.54  \\
  &  16165.032 & 7.71 & 7.62 & 7.71 & 7.72 & 7.78 & ... &  7.71 & 7.72 & 7.77 & 7.65 & 7.76  \\
$\langle A(Fe) \rangle$ &  & $7.61 \pm 0.07$ & $7.53 \pm 0.08$ & $7.60 \pm 0.08$ & $7.58 \pm 0.08$ & $7.62 \pm 0.09$ & $7.49 \pm 0.04$ & $7.61 \pm 0.09$ & $7.63 \pm 0.06$ & $7.61 \pm 0.10$ & $7.62 \pm 0.06$ & $7.64 \pm 0.07$  \\
Na I &  16373.853 & 6.67 &  6.67 &  6.62 &  6.55 &  6.65 &  6.49 &  6.60 &  6.67 &  6.64 &  6.61 &  6.62  \\
&  16388.858 & 6.74 &  6.73 &  6.66 &  6.62 &  6.72 &  6.55 &  6.65 &  ...  &  6.67 &  6.72 &  6.67  \\
$\langle A(Na) \rangle$ & & $6.71 \pm 0.04$ & $6.70 \pm 0.03$ & $6.64 \pm 0.02$ & $6.59 \pm 0.04$ & $6.68 \pm 0.03$ & $6.52 \pm 0.03$ & $6.63 \pm 0.03$ & $6.67$  & $6.66 \pm 0.02$ & $6.67 \pm 0.05$ & $6.65 \pm 0.03$ \\
Mg I &  15740.716 & 7.75 & 7.60 & 7.67 & 7.61 & 7.72 & 7.51 & 7.63 & 7.63 & 7.70 & 7.57 & 7.68  \\
  &  15748.9   & 7.72 & 7.63 & 7.63 & 7.52 & 7.65 & 7.51 & ... &  7.61 & 7.57 & 7.55 & 7.68  \\
  &  15765.8   & 7.69 & 7.60 & 7.69 & 7.60 & 7.77 & 7.60 & 7.70 & 7.76 & 7.70 & 7.50 & 7.59  \\
  &  15879.5   & 7.55 & 7.74 & 7.58 & 7.55 & 7.54 & 7.68 & 7.56 & 7.60 & 7.75 & 7.54 & 7.78  \\
  &  15886.2   & 7.80 & 7.81 & 7.84 & 7.78 & ... &  7.78 & ... &  ... &  7.86 & 7.73 & 7.82  \\
  &  15954.477 & 7.71 & 7.72 & 7.75 & 7.68 & 7.77 & 7.63 & ... &  7.67 & 7.74 & 7.59 & 7.66  \\
$\langle A(Mg) \rangle$ &  & $7.70 \pm 0.08$ & $7.68 \pm 0.08$ & $7.69 \pm 0.08$ & $7.62 \pm 0.09$ & $7.69 \pm 0.09$ & $7.62 \pm 0.10$ & $7.63 \pm 0.06$ & $7.65 \pm 0.06$ & $7.72 \pm 0.09$ & $7.58 \pm 0.07$ & $7.70 \pm 0.07$  \\
Al I &  16718.957 & 6.77 & 6.58 & 6.76 & 6.62 & ... &  6.53 & 6.70 & 6.68 & 6.71 & 6.65 & 6.72  \\
  &  16763.359 & 6.72 & 6.58 & 6.66 & 6.50 & 6.49 & 6.45 & 6.60 & 6.55 & 6.65 & 6.61 & 6.64  \\
$\langle A(Al) \rangle$ &  & $6.75 \pm 0.02$ & $6.58 \pm 0.00$ & $6.71 \pm 0.05$ & $6.56 \pm 0.06$ & $6.49$ & $6.49 \pm 0.04$ & $6.65 \pm 0.05$ & $6.61 \pm 0.06$ & $6.68 \pm 0.03$ & $6.63 \pm 0.02$ & $6.68 \pm 0.04$  \\
Si I &  15361.161 & 7.68 & 7.75 & 7.78 & 7.75 & 7.68 & 7.73 & 7.59 & 7.58 & 7.78 & 7.74 & 7.80  \\
  &  15376.831 & 7.89 & 7.83 & 7.83 & 7.85 & 7.80 & 7.75 & 7.84 & 7.78 & 7.86 & 7.87 & 7.92  \\
  &  16060.009 & 7.56 & 7.48 & 7.45 & 7.45 & 7.56 & 7.45 & 7.47 & 7.44 & 7.57 & 7.50 & 7.61  \\
  &  16094.787 & 7.79 & 7.60 & 7.71 & 7.61 & 7.68 & 7.48 & 7.63 & 7.70 & 7.76 & 7.80 & 7.85  \\
  &  16215.67  &  7.73 & 7.52 & 7.71 & 7.58 & 7.64 & 7.48 & 7.60 & 7.73 & 7.74 & 7.60 & 7.67  \\
  &  16680.77  &  7.70 & 7.63 & 7.78 & 7.61 & 7.73 & 7.55 & 7.63 & 7.73 & 7.77 & 7.70 & 7.73  \\
  &  16828.159 & ... &  7.61 & 7.78 & 7.61 & 7.65 & 7.52 & 7.55 & 7.65 & 7.75 & 7.61 & 7.67  \\
$\langle A(Si) \rangle$ &  & $7.73 \pm 0.10$ & $7.63 \pm 0.11$ & $7.72 \pm 0.12$ & $7.64 \pm 0.12$ & $7.68 \pm 0.07$ & $7.57 \pm 0.11$ & $7.62 \pm 0.10$ & $7.66 \pm 0.11$ & $7.75 \pm 0.08$ & $7.69 \pm 0.12$ & $7.75 \pm 0.10$  \\
K I & 15163.07 &  5.04 & 5.10 & 5.08 & 4.92 & 4.81 & 5.01 & 4.84 & 4.87 & ... &  5.07 & 5.10  \\
  &  15168.38 &  5.04 & 5.04 & 5.03 & 4.88 & 4.90 & 5.01 & 4.95 & 4.90 & 4.95 & 4.93 & 5.04  \\
$\langle A(K) \rangle$ &  & $5.04 \pm 0.00$ & $5.07 \pm 0.03$ & $5.06 \pm 0.02$ & $4.90 \pm 0.02$ & $4.86 \pm 0.05$ & $5.01 \pm 0.00$ & $4.90 \pm 0.06$ & $4.89 \pm 0.02$ & $4.95 $ & $5.00 \pm 0.07$ & $5.07 \pm 0.03$  \\
Ca I &  16150.763 & 6.55 & 6.33 & 6.43 & 6.42 & 6.47 & 6.30 & 6.40 & 6.50 & 6.57 & 6.42 & 6.39  \\
  &  16155.236 & 6.61 & 6.40 & 6.52 & 6.45 & 6.51 & 6.36 & 6.50 & 6.51 & 6.59 & 6.50 & 6.45  \\
  &  16157.364 & ... &  6.36 & 6.47 & 6.38 & 6.50 & 6.38 & 6.45 & 6.45 & 6.58 & 6.50 & 6.47  \\
$\langle A(Ca) \rangle$ &  & $6.58 \pm 0.03$ & $6.36 \pm 0.03$ & $6.47 \pm 0.04$ & $6.42 \pm 0.03$ & $6.49 \pm 0.02$ & $6.35 \pm 0.03$ & $6.45 \pm 0.04$ & $6.49 \pm 0.03$ & $6.58 \pm 0.01$ & $6.47 \pm 0.04$ & $6.44 \pm 0.03$  \\
Ti I &  15543.756 & 5.22 & 4.92 & 5.01 & 4.90 & 5.09 & 4.77 & 4.98 & 5.09 & 5.08 & 5.11 & 5.04  \\
  &  15602.842 & 5.28 & 5.02 & 5.17 & 4.94 & 5.23 & 5.01 & 5.20 & 5.25 & 5.14 & 5.28 & 5.13  \\
  &  15698.979 & 5.16 & 4.94 & 5.03 & 4.80 & 4.98 & 4.79 & 4.92 & ... & 5.13 & 4.95 & 5.08  \\
  &  15715.573 & 5.09 & 4.83 & ...  & 4.68 & 4.96 & 4.75 & 4.92 & 4.92 & 4.88 & 4.95 & 4.95  \\
  &  16635.161 & 5.19 & ...  & ...  & 4.83 & 4.84 & ...  & ...  & ...  & 4.92 & ...  & 4.93  \\
$\langle A(Ti) \rangle$ &  & $ 5.19 \pm 0.06$ & $4.93 \pm 0.07$ & $5.07 \pm 0.07$ & $4.83 \pm 0.09$ & $5.02 \pm 0.13$ & $4.83 \pm 0.10$ & $5.01 \pm 0.12$ & $5.03 \pm 0.13$ & $5.03 \pm 0.11$ & $5.07 \pm 0.14$ & $5.03 \pm 0.08$  \\
V I & 5924.769 &  4.08 & 4.01 & 4.15 & 3.98 & 4.10 & 4.11 & 4.19 & 4.10 & 4.13 & 4.09 & 4.00  \\
$\langle A(V) \rangle$ &  & 4.08 & 4.01 & 4.15 & 3.98 & 4.10 & 4.11 & 4.19 & 4.10 & 4.13 & 4.09 & 4.00  \\
Cr I &  15680.063 & 5.94 & 5.76 & 5.76 & 5.62 & 5.86 & 5.69 & 5.67 & 5.71 & 5.85 & 5.74 & 5.78  \\
$\langle A(Cr) \rangle$ &  & 5.94 & 5.76 & 5.76 & 5.62 & 5.86 & 5.69 & 5.67 & 5.71 & 5.85 & 5.74 & 5.78  \\
Mn I &  15159.  & 5.70 & 5.68 & 5.65 & 5.56 & 5.50 & 5.56 & 5.63 & 5.60 & 5.61 & 5.64 & 5.58  \\
  &  15217.  & 5.55 & ... &  5.48 & 5.35 & 5.42 & 5.33 & 5.40 & 5.43 & 5.45 & 5.56 & 5.47  \\
  &  15262.  & 5.70 & 5.59 & 5.67 & 5.50 & 5.55 & 5.48 & 5.53 & 5.50 & 5.61 & 5.68 & 5.70  \\
$\langle A(Mn) \rangle$ &  & $5.65 \pm 0.07$ & $5.63 \pm 0.04$ & $5.60 \pm 0.08$ & $5.47 \pm 0.09$ & $5.49 \pm 0.05$ & $5.46 \pm 0.09$ & $5.52 \pm 0.09$ & $5.51 \pm 0.07$ & $5.56 \pm 0.07$ & $5.63 \pm 0.05$ & $5.58 \pm 0.09$  \\
Co I &  16757.7 & 5.11 & 5.06 & 5.12 & 4.94 & 5.10 & 4.97 & 5.08 & 5.15 & 5.07 & 5.07 & 5.04  \\
$\langle A(Co) \rangle$ &  & 5.11 & 5.06 & 5.12 & 4.94 & 5.10 & 4.97 & 5.08 & 5.15 & 5.07 & 5.07 & 5.04  \\
Ni I &  15605.68 &  6.58 & 6.42 & 6.50 & 6.36 & 6.43 & 6.34 & 6.40 & 6.40 & 6.49 & 6.41 & 6.50  \\
  &  15632.654 & 6.56 & 6.38 & 6.45 & 6.41 & 6.48 & 6.30 & 6.42 & 6.45 & 6.54 & 6.39 & 6.52  \\
  &  16584.44 &  6.56 & 6.39 & 6.50 & 6.45 & 6.49 & 6.44 & 6.57 & 6.48 & 6.59 & 6.42 & 6.49  \\
  &  16589.30 &  6.43 & 6.34 & 6.51 & 6.41 & 6.36 & 6.37 & 6.47 & 6.31 & 6.51 & 6.47 & 6.46  \\
  &  16673.711 & ... &  6.28 & 6.30 & 6.37 & 6.27 & 6.30 & 6.37 & 6.31 & 6.32 & 6.35 & 6.29  \\
  &  16815.471 & 6.51 & 6.46 & 6.40 & 6.28 & 6.37 & 6.39 & 6.40 & 6.37 & 6.32 & 6.51 & 6.43  \\
  &  16818.76 &  ... &  6.54 & 6.58 & 6.43 & 6.51 & 6.52 & 6.48 & 6.34 & 6.56 & ... &  6.55  \\
$\langle A(Ni) \rangle$ &  & $6.53 \pm 0.05$ & $6.40 \pm 0.08$ & $6.46 \pm 0.08$ & $6.39 \pm 0.05$ & $6.41 \pm 0.08$ & $6.38 \pm 0.07$ & $6.45 \pm 0.06$ & $6.38 \pm 0.06$ & $6.48 \pm 0.10$ & $6.43 \pm 0.05$ & $6.46 \pm 0.08$  \\
Ce II & 15784.786 & 1.86 & 1.79 & 1.72 & 1.79 & 1.82 & 1.82 & 1.80 & ... &  1.82 & 1.84 & 1.79  \\  
  &  15958.39 &  1.90 & ... &  ... &  1.89 & ... &  1.84 & ... &  2.05 & ... &  1.72 & 1.87  \\
  &  15977.12 &  1.83 & ... &  1.81 & ... &  ... &  ... &  ... &  ... &  1.75 & 1.72 & ...   \\
  &  16327.32 &  ... &  ... &  ... &  ... &  ... &  ... &  1.70 & ... &  ... &  ... &  ...   \\
  &  16376.46 &  1.78 & 1.91 & 1.91 & 1.85 & 1.93 & 1.80 & 1.80 & 1.90 & ... &  ... &  1.96  \\
  &  16595.233 & 1.97 & 1.75 & 1.82 & ... &  1.84 & ... &  1.77 & 1.92 & 1.74 & 1.90 & 1.69  \\
  &  16722.6 &   ... &  1.98 & ... &  2.03 & 1.96 & 1.92 & ... &  2.06 & ... &  ... &  2.00  \\
$\langle A(Ce) \rangle$ &  & $1.87 \pm 0.06$ & $1.86 \pm 0.09$ & $1.81 \pm 0.07$ & $1.89 \pm 0.09$ & $1.89 \pm 0.06$ & $1.84 \pm 0.05$ & $1.77 \pm 0.04$ & $1.98 \pm 0.07$ & $1.77 \pm 0.04$ & $1.80 \pm 0.08$ & $1.86 \pm 0.11$
\end{longtable}
\end{landscape}
}
\clearpage
\twocolumn
\section{Comparisons with Previous Results} \label{sec:compar}
\subsection{Stellar Parameters}
As discussed previously, the derived stellar parameters in this study relied on photometric calibrations for the derivation of the effective temperature, and fundamental relations for the derivation of the surface gravity. In the following, we compare our parameters with those obtained from high-resolution spectroscopy in the literature.

Several of the more recent studies of the open cluster NGC 6705 presented results from the GES, such as, \citet[][25 stars]{Cantat2014}, \citet[][27 stars]{Tautvaiviense2015},  \citet[][15 stars]{Magrini2017}, and \citet[][71 stars]{magrini2021A&A...651A..84M}, who used effective temperatures and surface gravities from different survey data releases: GES-viDR1-final, GESviDR2Final, GES-IDR4, and GES-iDR6, respectively. In addition to these studies, there are also results for the NGC 6705 stars studied here in GES-DR5 (latest data release). Another recent study in the literature by \citet[][]{Casamiquela2018AA...610A..66C}, analyzed eight stellar members of NGC 6705 observed by the OCCASO survey \citep{casamiquela2016MNRAS.458.3150C}.

\begin{figure*}
\centering
\includegraphics[width=0.325\textwidth]{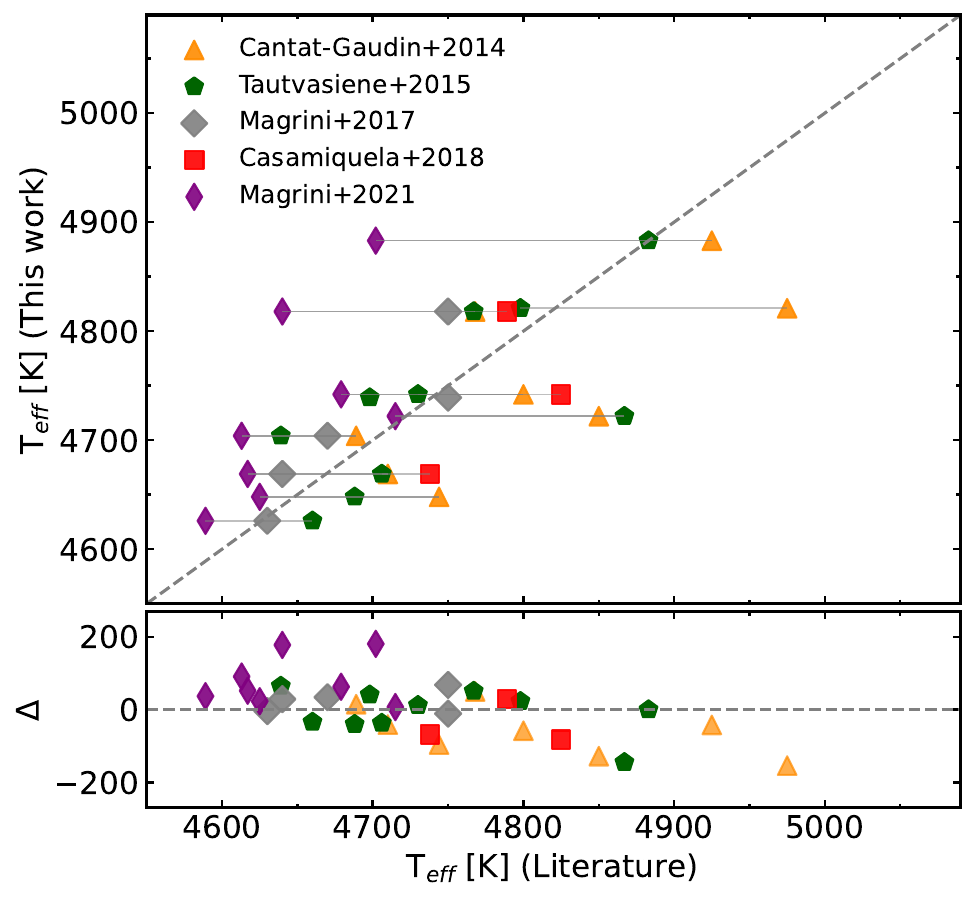}
\includegraphics[width=0.325\textwidth]{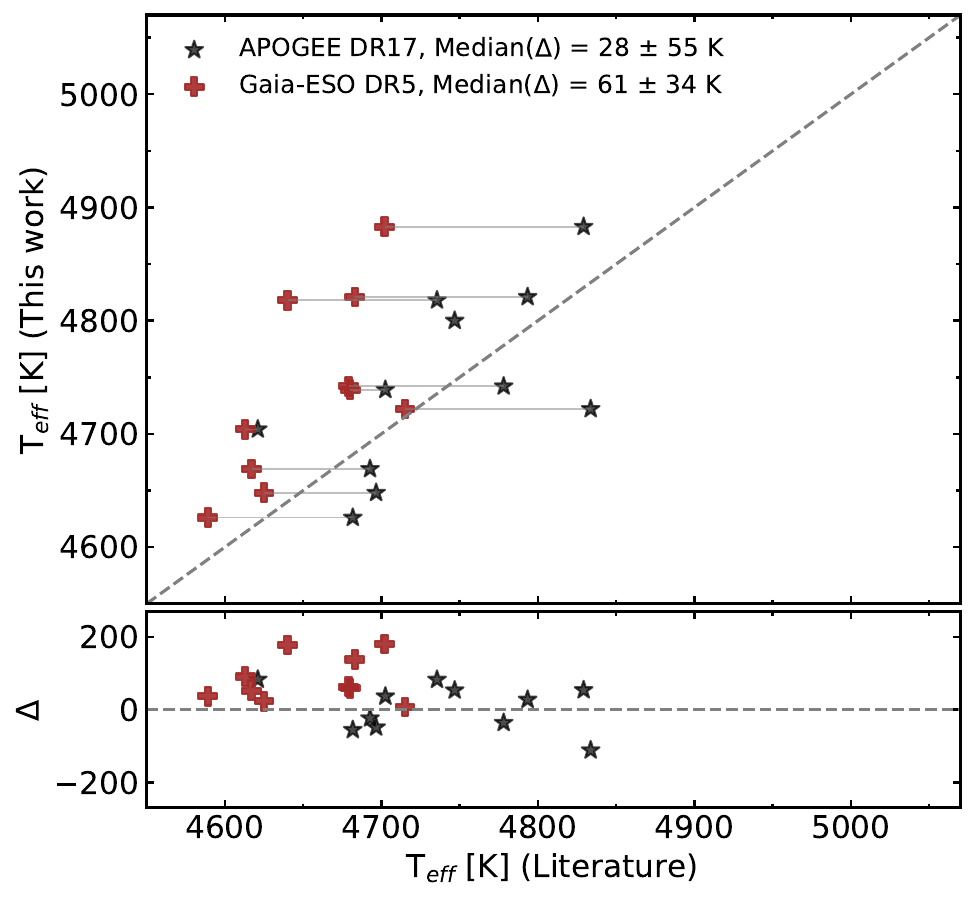}\\
\includegraphics[width=0.325\textwidth]{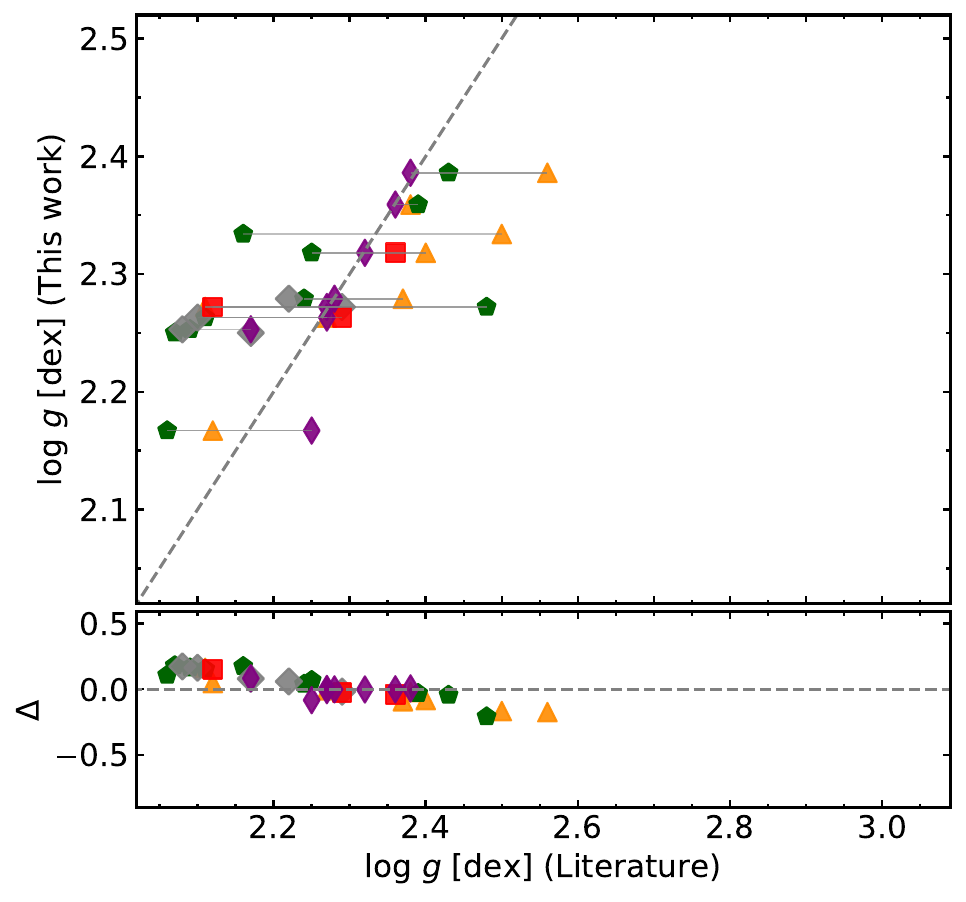}
\includegraphics[width=0.325\textwidth]{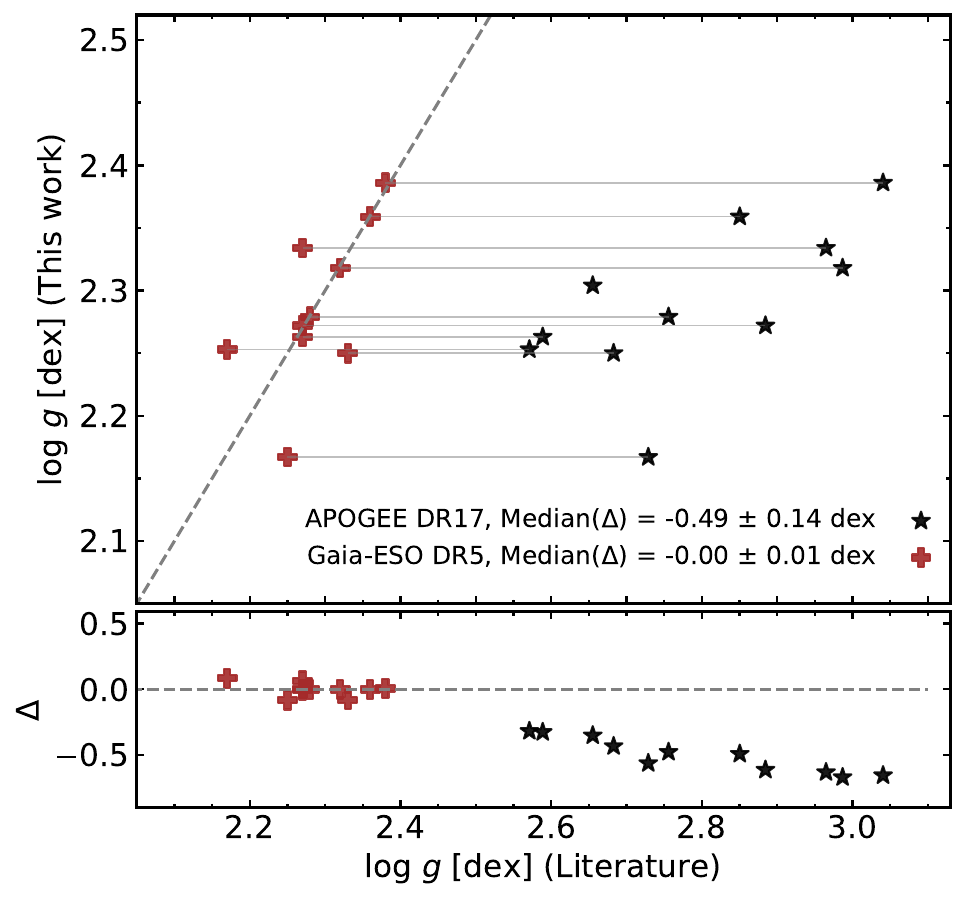}
\caption{Comparison of the effective temperatures (top panels) and surface gravities (bottom panels) derived in this work with the literature results. The orange triangles show the comparison with Cantat-Gaudin et al. \citeyear{Cantat2014}, the green pentagons with  Tautvai{\v s}ien{\.e} et al.  \citeyear{Tautvaiviense2015}, gray diamonds  with Magrini et al. \citeyear{Magrini2017}, the red squares with Casamiquela et al. \citeyear{Casamiquela2018AA...610A..66C}, narrow purple diamonds with Magrini et al. \citeyear{magrini2021A&A...651A..84M}, black stars with APOGEE DR17 (non-calibrated), and the brown cross with GES DR5. Results for the same star are connected by the gray solid lines. The gray dashed lines represent equality. The lower panels show the differences, $\Delta$, ``This work - Others" for effective temperatures and surface gravities, respectively.}
\label{fig:comp_par}
\end{figure*}

A comparison of the effective temperatures for stars in common between our work and the studies mentioned above is shown in the top left panel of Figure \ref{fig:comp_par}, while the top right panel shows the ASPCAP pipeline \citep{Garcia2016} results from APOGEE DR17 and  GES-DR5: residual differences as a function of the literature $T_{\rm eff}$ are shown at the bottom of the two panels.
The median differences between the effective temperatures in ``This work - Others" ($\pm$MAD) are as follows: $-42\pm57 $ K for \cite{Cantat2014} (9 stars); $ 6\pm41 $ K for \cite{Tautvaiviense2015} (10 stars); $ 29 \pm 33 $ K for \cite{Magrini2017} (5 stars); $ -69 \pm 14 $ K for \cite{Casamiquela2018AA...610A..66C} (3 stars); $ 57.5 \pm 34.0 $ for \cite{magrini2021A&A...651A..84M} (8 stars). 
The comparison with results GES-DR5 finds $<$$\Delta T_{\rm eff}$$>$ = $ 61\pm34 $ K (10 stars) and with ASPCAP APOGEE DR17 (uncalibrated) is $28\pm55$ K (11 stars). In general, all analyses yield consistent values of T$\rm _{eff}$ within typical uncertainties in effective temperature scales.

The two bottom panels of Figure \ref{fig:comp_par} are equivalent to the top panels but the comparison is for surface gravities. The median differences ($\pm$ MAD) in log g between ``This Work - Others" (left bottom panel) are, respectively: $-0.06\pm0.05$ dex for \citep{Cantat2014}; $0.09\pm0.08$ dex for \citep{Tautvaiviense2015}, $0.08 \pm 0.08$ dex for \citep{Magrini2017}; for $-0.03 \pm 0.01$ dex for \citep{Casamiquela2018AA...610A..66C}. Although there is overall good agreement in these median log g differences, which are well within the expected uncertainties for spectroscopically-determined values of log g (e.g., $\sim$0.2 dex), there are clear trends in the results, as can be seen from $\Delta$log g as a function of log g shown in the bottom subpanel. We generally find a smaller range in log $g$ values for the comparison sample than the literature.
However, our log g values are in excellent agreement with GES-DR5:  $\langle\Delta $log g$\rangle =-0.0\pm0.01$ dex. APOGEE surface gravity results on the other hand, are known to have significant offsets (both for red giants and dwarfs) and this is clear from the bottom right panel of Figure \ref{fig:comp_par}. The range in log $g$ from APOGEE DR17 (ASPCAP derived values) varies roughly between 2.6 and 3.1 dex and the median differences in log $g$ values show systematics: $-0.49 \pm 0.14$ dex. APOGEE also provides post-calibrated log $g$s and in this case the median log $g$ differences are improved: $-0.31 \pm 0.14$ dex.

\begin{table}
\centering
\caption{Mean NGC 6705 Abundances}
\begin{tabular}{lcccc}
\hline \hline
Element & $\langle A(X) \rangle$ & $\langle [X/H] \rangle$ & $STD$ & $\sigma (X)$ \\ \hline
C  & 8.44 & $-$0.02 & 0.07 & 0.057 \\
N  & 8.47 & 0.64 & 0.05 & 0.085 \\
O  & 8.76 & 0.07 & 0.03 & 0.132 \\
Fe & 7.59 & 0.13 & 0.04 & 0.035 \\
Na & 6.65 & 0.43 & 0.05 & 0.035 \\
Mg & 7.66 & 0.11 & 0.04 & 0.078 \\
Al & 6.62 & 0.19 & 0.08 & 0.055 \\
Si & 7.68 & 0.17 & 0.06 & 0.055 \\
K  & 4.98 & $-$0.09 & 0.08 & 0.053 \\
Ca & 6.46 & 0.16 & 0.07 & 0.058 \\
Ti & 5.01 & 0.04 & 0.10 & 0.103 \\
V  & 4.06 & 0.19 & 0.06 & 0.051 \\
Cr & 5.76 & 0.14 & 0.09 & 0.032 \\
Mn & 5.55 & 0.13 & 0.07 & 0.062 \\
Co & 5.06 & 0.12 & 0.06 & 0.067 \\
Ni & 6.43 & 0.23 & 0.05 & 0.044 \\
Ce & 1.85 & 0.27 & 0.06 & 0.101 \\
\hline
\end{tabular}
\label{tab:mean_abu}
\end{table}

\subsection{Metallicities \& Other Elemental Abundances} \label{sec:metal}
There are several results for the metallicity of NGC 6705 in the literature. 
Metallicities derived spectroscopically, with values between +0.02 dex and +0.24 dex, are reported in studies by \citet{Gonzalez2000,Magrini2014,Tautvaiviense2015,Magrini2017,Casamiquela2018AA...610A..66C,magrini2021A&A...651A..84M,casamiquela2021A&A...652A..25C}. 
According to our results based on near-infrared spectroscopic analysis of a sample of eleven red giant stars, the mean metallicity of NGC 6705 is $\rm \langle [Fe/H]\rangle = +0.13 \pm 0.04$ dex. 

\begin{figure}
\centering
\includegraphics[width=0.4\textwidth]{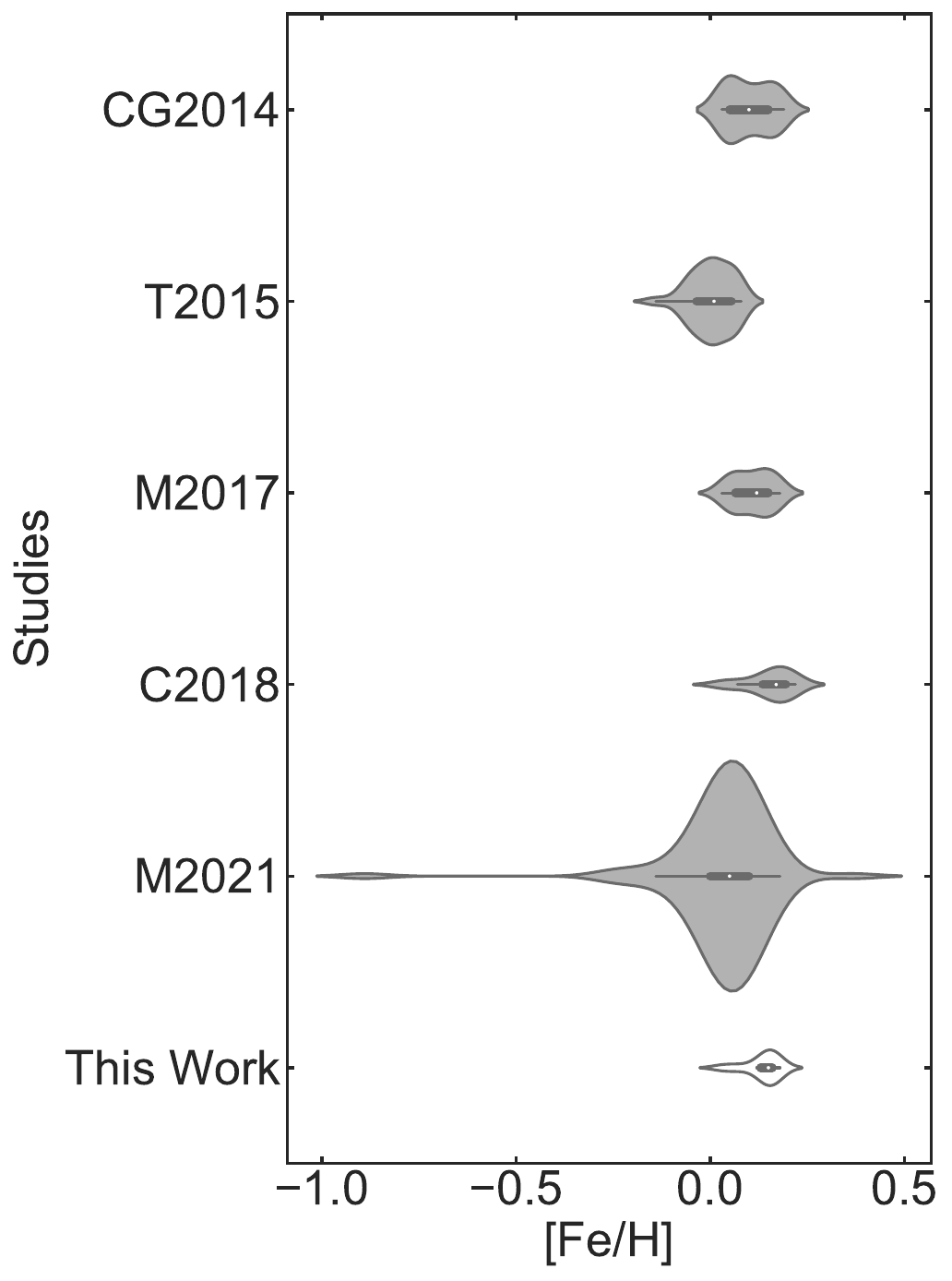}
\caption{Metallicity distribution of NGC 6705 stars. The white distribution represents our [Fe/H] results, while the gray sequences show the literature [Fe/H] results: \citet[][CG2014]{Cantat2014}, \citet[][T2015]{Tautvaiviense2015}, \citet[][M2017]{Magrini2017}, \citet[][C2018]{Casamiquela2018AA...610A..66C}, and \citet[][M2021]{magrini2021A&A...651A..84M}. White dots in the distribution indicate the median, while the thick bar represents the interquartile range, and the thin bar shows the 95\% confidence interval. Wider regions of the distribution represent a higher probability that a star will have that [Fe/H] value.
\label{fig:vio_feh}}
\end{figure}

Figure \ref{fig:vio_feh} shows violin distributions of the metallicity of NGC 6705 determined both in this study (white distribution) and other studies (gray distributions). The metallicity diagrams from this work, as well as \cite{Cantat2014}, \cite{Tautvaiviense2015}, \cite{Magrini2017}, \cite{Casamiquela2018AA...610A..66C}, and \cite{magrini2021A&A...651A..84M} are shown from top to bottom in chronological order. In general, our metallicity determination shows excellent agreement with \cite{Magrini2017} and \cite{Casamiquela2018AA...610A..66C} for their samples of 15 and 8 stars, with reported metallicities of [Fe/H] = $+0.12\pm0.05$ dex and [Fe/H] = $+0.17\pm0.04$ dex, respectively, or mean metallicity differences of +0.01 and $-0.04$ dex, respectively. However, the mean metallicity value determined here is somewhat larger than those reported by \cite{Tautvaiviense2015} in a sample of 27 stars ($\rm [Fe/H]=0.00\pm0.05$ dex) and \cite{magrini2021A&A...651A..84M} using 71 stars ($0.02\pm0.05$ dex), with the latter study having used both high-resolution UVES and lower resolution GIRAFFE data.

In addition, an investigation into the differences between the metallicities of ``This work - Other" ($\pm$MAD) for stars in common between the studies, finds good agreement in some cases and systematic differences in others. There are no significant systematic differences when comparing with  \cite{Casamiquela2018AA...610A..66C}, \cite{Magrini2017}, and \cite{Cantat2014}, with median differences of $0.00\pm0.01$ dex, $0.03\pm0.02$ dex, and $0.03\pm0.04$ dex, respectively. The comparison with results from APOGEE DR17 is also very good: $0.04\pm0.03$ dex. On the other hand, there are larger differences between the metallicities of \cite{Tautvaiviense2015}, \cite{magrini2021A&A...651A..84M}, and GES DR5, for ten, eight and ten stars, with our metallicities being higher than theirs by $0.13\pm0.04$ dex, $0.12\pm0.04$ dex, and $0.12\pm0.02$ dex, respectively.

\begin{table*}
\centering
\caption{Median Abundance differences ``This Work - Other"}
\begin{tabular}{lccccccc}
\hline \hline
$[X/Fe]$ & APOGEE DR17 & GES DR5 & CG2014 & T2015 & M2017 & C2018 & SS2022 \\
 & ($\# 11$) & ($\# 10$) & ($\# 8$) & ($\# 10$) & ($\# 5$) & ($\# 3$) & ($\# 9$) \\ \hline
C   & -0.18 $\pm$ 0.03 & -0.06 $\pm$ 0.04 &  ...              &  -0.06 $\pm$ 0.03 &  ...             &  ...              &  ...              \\
N   &  0.06 $\pm$ 0.03 & -0.05 $\pm$ 0.06 &  ...              &  -0.12 $\pm$ 0.03 &  ...             &  ...              &  ...              \\
O   &  0.03 $\pm$ 0.02 & -0.20 $\pm$ 0.03 &  ...              &  -0.18 $\pm$ 0.08 & -0.03 $\pm$ 0.04 &  -0.25 $\pm$ 0.07 &  ...             \\
Na  &  0.11 $\pm$ 0.05  &  0.03 $\pm$ 0.10  &  -0.26 $\pm$ 0.03  &  ...  &  ...  &  ...  &  ...              \\
Mg  &  0.11 $\pm$ 0.03 & -0.19 $\pm$ 0.06 &  -0.16 $\pm$ 0.04 &  ...              & -0.10 $\pm$ 0.04 &  -0.14 $\pm$ 0.11 &  ...              \\
Si  &  0.01 $\pm$ 0.03 & -0.04 $\pm$ 0.07 &   0.03 $\pm$ 0.02 &  ...              &  0.04 $\pm$ 0.04 &  -0.05 $\pm$ 0.06 &  ...              \\
Ca  &  0.04 $\pm$ 0.02 &  0.12 $\pm$ 0.09 &  ...              &  ...              &  0.11 $\pm$ 0.03 &  -0.05 $\pm$ 0.10 &  ...              \\
Al  & -0.00 $\pm$ 0.04 & -0.11 $\pm$ 0.05 &  -0.15 $\pm$ 0.06 &  ...              &  ...             &  ...              &  ...              \\
K   & -0.04 $\pm$ 0.06 & ...              &  ...              &  ...              &  ...             &  ...              &  ...              \\
Ti  &  -0.06 $\pm$ 0.05  &  -0.01 $\pm$ 0.10  &  ...  &  ...  &  0.05 $\pm$ 0.00  &  -0.11 $\pm$ 0.01  &  ...             \\
V   &  0.26 $\pm$ 0.03 &  0.05 $\pm$ 0.04 &  ...              &  ...              &  0.02 $\pm$ 0.01 &  ...              &  ...              \\
Cr  &  0.03 $\pm$ 0.06 &  0.09 $\pm$ 0.10 &  ...              &  ...              &  0.15 $\pm$ 0.04 &  ...              &  ...              \\
Mn  &  0.01 $\pm$ 0.06 &  0.07 $\pm$ 0.06 &  ...              &  ...              &  ...             &  ...              &  ...              \\
Co  & -0.03 $\pm$ 0.04 & -0.05 $\pm$ 0.05 &  ...              &  ...              &  ...             &  ...              &  ...              \\
Ni  &  0.10 $\pm$ 0.04 &  0.12 $\pm$ 0.05 &  ...              &  ...              &  0.12 $\pm$ 0.03 &  ...              &  ...              \\
Ce  &  0.01 $\pm$ 0.07 &  0.05 $\pm$ 0.07 &  ...              &  ...              &  ...             &  ...              &  -0.07 $\pm$ 0.02 \\
\hline
\end{tabular}
\begin{tablenotes}
\item \textbf{Notes}: Our results are compared with: CG2014: \cite{Cantat2014}; T2015: \cite{Tautvaiviense2015}; M2017: \cite{Magrini2017}; C2018: \cite{Casamiquela2018AA...610A..66C}; SS2022: \cite{sales2022ApJ...926..154S}
\end{tablenotes}
\label{tab:mean_xfe}
\end{table*}

Besides metallicities, which are discussed above, we summarize in Table \ref{tab:mean_xfe} the comparisons between the abundances of the other elements studied here with literature values, again exemplified as the median abundance differences ($\pm$ MAD) [X/Fe] for ``This Work - Other Work" for stars in common with the studies of \cite{Cantat2014}, \cite{Tautvaiviense2015}, \cite{Magrini2017}, \cite{Casamiquela2018AA...610A..66C}, and the surveys APOGEE DR17 and GES-DR5. Most of the systematic differences between the results are less than or equal to 0.1 dex and this is not surprising given the different methodologies adopted in the various studies, but there are some cases with more significant discrepancies, such as, for example, oxygen having a median difference of $-0.25$ dex for \cite{Casamiquela2018AA...610A..66C} and $-0.20$ dex for GES-DR5, magnesium being different by $-0.19$ dex also for GES-DR5, or sodium having a $-$0.26 dex offset in comparison with \cite{Cantat2014}. In addition, the MAD values for the majority of the cases are also typically small ($<$ 0.05 dex), with only one being larger than 0.1 dex.

\section{Discussion} \label{sec:res_disc}

\subsection{C, N, O, Na, Al, and Mixing in the Red-giants of NGC 6705} \label{sec:mixing}
NGC 6705 provides an important astrophysical laboratory in which to probe red-giant mixing in RGB and RC stars, as the masses of this cluster's red giants are M$\sim$3.3 M$_{\odot}$, well above the mass limit for stars that undergo the He-core flash \citep[M$<$2.1 M$_{\odot}$ at solar metallicity, e.g.,][]{karakas2014PASA...31...30K}. 
As discussed in Section \ref{sec:analysis} and illustrated in Figure \ref{fig:path_evol}, the red giants sampled here are likely a mixture of RGB and RC stars, with the interiors of the RGB stars consisting of an inert He core surrounded by a H-burning shell, while the RC stars have evolved beyond the RGB and are powered by core-He burning.  Due to uncertainties in our derived values of $T_{\rm eff}$ and log $g$, assigning a classification to a red giant as either an RGB or RC star (without asteroseismic data) is uncertain, although based on their positions in the Kiel diagram in Figure \ref{fig:path_evol}, it appears that those red giants with $T_{\rm eff}>4750-4800$K likely belong to the RC, while the cooler ones are evolving up the RGB: this classification criterion results in our sample dividing into four RC stars and seven RGB stars. 

The luminosities of the RGB stars indicate that all have experienced the first dredge-up (as have the RC stars), which has contaminated their photospheres with matter that has undergone partial H-burning via the CN-cycle.  This contamination is revealed through the abundances of carbon, both $^{12}$C and $^{13}$C (although carbon-13 will not be discussed here) and $^{14}$N.  As a result of the first dredge-up, the surface $^{12}$C abundance will be lowered, while that of $^{14}$N will be increased significantly.  
As shown in Table \ref{tab:mean_abu}, the average carbon and nitrogen abundances are $\langle A(^{12}C)\rangle = 8.44 \pm 0.07$ and $\langle A(^{14}N)\rangle = 8.47 \pm 0.05$, respectively.  The solar abundances are A($^{12}$C)=8.46 and A($^{14}$N)=7.83, but with the average metallicity of NGC 6705 being [Fe/H]=+0.13, the average cluster red giant abundances of carbon and nitrogen with respect to iron are  [$^{12}$C/Fe]=$-0.16$ and [$^{14}$N/Fe]=+0.51, in qualitative agreement with that expected from first dredge-up. Theoretical models of the 1st dredge-up and thermohaline mixing by \cite{charbonnel2010A&A...522A..10C} predict [$^{12}$C/Fe]=$-0.21$ and [$^{14}$N/Fe]=+0.47 for stars with M=3 M$_{\odot}$ after dredge-up, which is in quantitative agreement with our results for the red giants in NGC 6705.

\begin{figure}
\centering
\includegraphics[width=0.5\textwidth]{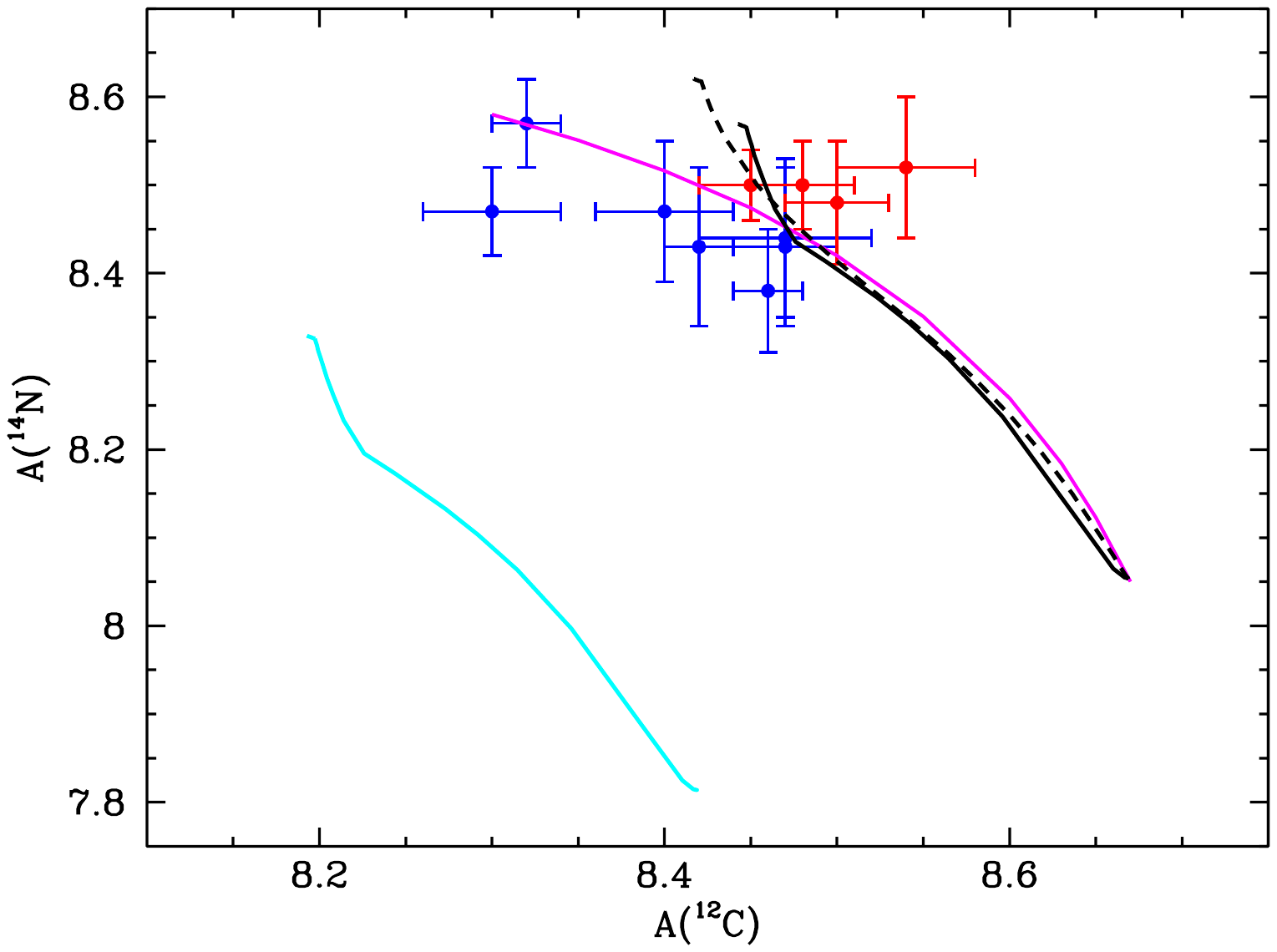}
\caption{Carbon-12 and nitrogen-14 abundances for the NGC 6705 red giants divided into RGB (blue symbols) and RC (red symbols) stars. The solid magenta curve is a ``mixing line'' defined by a constant sum of $^{12}$C and $^{14}$N nuclei.  A 3 M$_{\odot}$ solar-metallicity standard stellar model (i.e., no extra mixing) from Lagarde et al. (\citeyear{lagarde2012AA...543A.108L}) is shown as a cyan curve, while the black curves represent solar metallicity models simply shifted by +0.22 dex in the initial $^{12}$C and $^{14}$N abundances to mimic the metallicity of NGC 6705, with the standard model shown by the solid line and a model that includes rotational and thermohaline mixing shown by the dashed line.}
\label{fig:CN}
\end{figure}

Figure \ref{fig:CN} presents a different way to view the 1st dredge-up, with the $^{14}$N-abundance plotted versus the $^{12}$C-abundance, and the observed red giants in NGC 6705 are divided into RGB (blue symbols) and RC (red symbols) stars, respectively.  The smooth magenta curve represents a constant sum of the number abundances of carbon-12 and nitrogen-14, as the sum of these nuclei are conserved approximately during CN-cycle H-burning.  This curve represents schematically the 1st dredge-up as a mixing curve.  There are four stable nuclei involved in the CN-cycle: $^{12}$C, $^{13}$C, $^{14}$N, and $^{15}$N, of which carbon-13 and nitrogen-15 are considered minor species.   In pure equilibrium CN-cycle matter, the value of $^{12}$C/$^{13}$C can be as small as 3.5, which would shift the upper part of the mixing curve in Figure \ref{fig:CN} to lower $^{12}$C abundances by $\sim$0.1 dex, however, the expected ratio in 3.3 M$_{\odot}$ RGB stars is 18--20 \citep{lagarde2012AA...543A.108L,McCormick2023MNRAS.524.4418M}, which would have a negligible effect on the mixing curve.  Nitrogen-15 is an even more minor species, with typical 1st dredge-up values expected to be smaller than the solar value of $^{14}$N/$^{15}$N=272 \citep{wannier1991ApJ...380..593W}, so this isotope can be neglected from our discussion. 

The sum of N($^{12}$C) + N($^{14}$N) in Figure \ref{fig:CN} was taken as the average values from the RGB plus RC stars, and the initial individual carbon-12 and nitrogen-14 abundances were set assuming an initial solar ratio of N(C)/N(N)=4.1 \citep{Grevesse2007}. These two constraints lead to initial carbon and nitrogen abundances for NGC 6705 of A($^{12}$C)=8.67 and A($^{14}$N)=8.05 and, with these initial abundance values, the mixing curve passes through the NGC 6705 RGB stars. One point to note from Figure \ref{fig:CN} is that the four RC stars, while having $^{14}$N abundances that are very similar to the RGB stars, exhibit $^{12}$C abundances that are slightly larger than the RGB sample.  The average abundances of the two groups are $\langle$A($^{12}$C)$\rangle$=8.41$\pm$0.07 and $\langle$A($^{14}$N)$\rangle$=8.46$\pm$0.05 for the RGB stars and $\langle$A($^{12}$C)$\rangle$=8.49$\pm$0.03 and $\langle$A($^{14}$N)$\rangle$=8.50$\pm$0.01 for the RC stars, resulting in linear values for the C/N abundance ratios of 0.92$\pm$0.22 and 0.99$\pm$0.07 for the RGB and RC stars, respectively.  A Kolmogorov–Smirnov test of the C/N ratios in the RGB and RC stars finds that they can be represented by a single population in C/N.  

Models by \cite{lagarde2012AA...543A.108L} for a 3 M$_{\odot}$ solar-metallicity star are also shown in Figure \ref{fig:CN} as a comparison to the simple mixing curve.  As \cite{lagarde2012AA...543A.108L} 
only presented solar-metallicity, or lower, models, we show their solar-metallicity model as the continuous cyan curve, which begins at an initial abundance of A($^{12}$C)=8.43 and A($^{14}$N)=7.83 and evolves from there.  Since our discussion from above indicates that NGC 6705 is metal-rich relative to the Sun, the black curves in Figure \ref{fig:CN} represent solar-metallicity models from \cite{lagarde2012AA...543A.108L} in which the initial $^{12}$C and $^{14}$N abundances are increased by +0.22 dex; the solid black curve is the standard model, while the dashed curve represents the model that includes rotational and thermohaline mixing.  A quantitative comparison of models with observationally-derived abundance would demand consistent models, however this straightforward test indicates that the 11 red giant members of NGC 6705 analyzed here, at M$\sim$3.3 M$_{\odot}$, display unremarkable C and N abundances when compared to stellar models.     

\begin{figure*}
    \centering
    \includegraphics[width=1.0\textwidth]{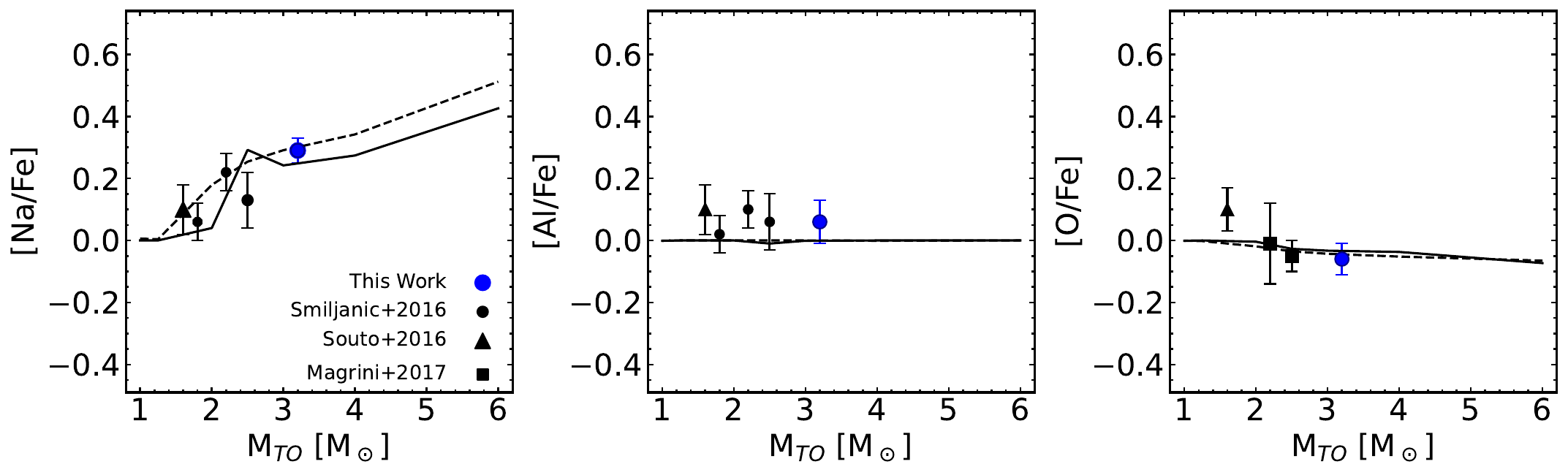}
    \caption{Mean cluster abundances of Na, Al, and O versus cluster turn-off mass. The blue circle corresponds to the abundances of NGC 6705 determined in this study, and the black circles correspond to mean Na and Al abundances of NGC 4815, Berkeley 81, and Trumpler 20 derived for giant stars in Smiljanic et al. (\citeyear{smiljanic2016AA...589A.115S}). Mean abundances of O are taken from Magrini et al. (\citeyear{Magrini2017}) for NGC 4815 and Berkeley 81 (black squares). The black triangle correspond to mean Na, Al, and O abundances of NGC 2420 derived in Souto et al. (\citeyear{Souto2016}). The black solid and dashed lines correspond to the standard model and the model that includes rotational and thermohaline mixing, both from Lagarde et al. (\citeyear{lagarde2012AA...543A.108L}), respectively.}
    \label{fig:NaAlFe_mass}
\end{figure*}

Moving up the periodic table past C and N, we investigate additional elemental abundances that are potentially sensitive to red giant mixing in the mass range of M$\sim 3-4$ M$_{\odot}$ and focus on oxygen (as $^{16}$O), sodium, and aluminum.  The red giants studied here have a mass of $\sim$3.3 M$_\odot$ and exhibit a significant overabundance of $\rm [Na/Fe] = +0.29 \pm 0.04$ dex. This confirms that giant stars with masses greater than 3 M$_\odot$ can have an overabundance of sodium, providing a strong indication that the sodium overabundance in these stars is caused by internal evolutionary processes, as suggested by \cite{smiljanic2016AA...589A.115S}.  
More specifically, we examine the behaviors of [Na/Fe], [Al/Fe], and [O/Fe] as functions of stellar mass in Figure \ref{fig:NaAlFe_mass} and, in addition to NGC 6705, we consider the slightly more metal poor ($\rm [Fe/H]=-0.16$) open cluster NGC 2420 (M$_{TO}$ = 1.6 M$_\odot$) with Na, Al, and O abundances taken from \cite{Souto2016}, the open clusters NGC 4815 (M$_{TO}$ = 2.5 M$_\odot$), Berkeley 81 (M$_{TO}$ = 2.2 M$_\odot$), and Trumpler 20 (M$_{TO}$ = 1.8 M$_\odot$), for which Na and Al abundances and masses were taken from \cite{smiljanic2016AA...589A.115S}, while the oxygen abundances are from \cite{Magrini2017}; the models from \cite{lagarde2012AA...543A.108L}, both standard as well as those including rotation, are also shown.  
The left panel of Figure \ref{fig:NaAlFe_mass} reveals an overabundance of sodium which increases with increasing red-giant mass, as predicted by the models. 
In the case of Al (middle panel of Figure \ref{fig:NaAlFe_mass}), we observe that the mean Al abundance of NGC 6705 determined here is slightly enhanced ($\rm [Al/Fe] = +0.06 \pm 0.07$) but in agreement with the models within the uncertainties, with all clusters displaying a small (but not significant) offset from the models. In summary, we do not find that the [Al/Fe] abundance is enhanced in NGC 6705, unlike what was previously suggested in \cite{smiljanic2016AA...589A.115S} for which the [Al/Fe] overabundance was +0.3. 
The right panel of Figure \ref{fig:NaAlFe_mass} shows [O/Fe] as a function of mass.  According to the models of \cite{lagarde2012AA...543A.108L}, there is a small trend of decreasing oxygen with increasing stellar mass and our oxygen abundance for NGC 6705 is also in agreement with the models.  Overall, the observed abundances of $^{12}$C, $^{14}$N, $^{16}$O, Na, and Al in the NGC 6705 red giants provide a benchmark for models of red giant mixing in intermediate-mass stars of M$\sim$3--4 M$_{\odot}$ at near-solar metallicity.

\subsection{Is the young open cluster NGC 6705 $\alpha$-enhanced?} \label{sec:alpha}

As discussed in the introduction, one of the interesting recent results in the literature is the finding that there is a population of stars in the Galaxy that is young and enriched in [$\alpha$/Fe]; a result which was based on stellar ages obtained via CoRoT asteroseismology and chemical abundances taken from the APOGEE survey \citep{Chiappini2015}. 
\cite{Martig2015} also identified young $\alpha$-enhanced stars using independent age estimates inferred from Kepler asteroseismology. In simple terms, such population is unexpected because in principle a young star is formed from gas already enriched in Fe from SN Ia, having therefore a decreased [$\alpha$/Fe] ratio. However, these $\alpha$-enhanced stars that appear to be young may be products binary interactions/mergers \citep[e.g.,][]{izzard2018MNRAS.473.2984I,silvaaguirre2018MNRAS.475.5487S,hekker2019MNRAS.487.4343H,jofre2023AA...671A..21J}, being actually old and having the expected $\alpha$-Fe content for their age. Moreover,
\cite{miglio2021AA...645A..85M} identified a sample of 400 red giant stars from the Kepler field (having asteroseismic data) that belong to the thick disk ([$\rm \alpha/Fe]>+0.1$) and found that $\sim$5\% of stars on the RGB were overmassive given the estimated age of the thick disk stars of $\sim$11 Gyr (M $>$1.1 M$_{\odot}$).  This fraction of overmassive stars increased to $\sim$18\% for red clump giants and \cite{miglio2021AA...645A..85M} suggest that this result supports the idea that these stars increased their initial masses via interactions with a binary companion (either mass transfer or mergers) while evolving up the RGB. 

In this context, the results in the literature finding the members of the young open cluster NGC 6705 to be $\alpha$-enhanced is puzzling. \cite{Casamiquela2018AA...610A..66C} studied a sample with eight stars members of the open cluster NGC 6705 from high-resolution optical spectra and found that they were enriched in $\alpha$-elements with an average of $\langle [\alpha/Fe]\rangle=+0.11\pm0.06$. \cite{Tautvaiviense2015} also studied NGC 6705 but focused only on the analysis of the elements carbon, nitrogen, and oxygen, which are involved in H-burning. They found that the mean oxygen abundance in their sample of 27 red giants was mildly enhanced, with $\rm \langle [O/Fe] \rangle = 0.12\pm0.05$. \cite{Magrini2014,Magrini2017} also found evidence of enhancements in some of the $\alpha$-elements in this cluster.
Such results for an open cluster provided a possible connection with the young and $\alpha$-enhanced field stars in the Galaxy, which was examined in  \cite{Casamiquela2018AA...610A..66C}.

\begin{figure*}
\centering
\includegraphics[width=0.7\textwidth]{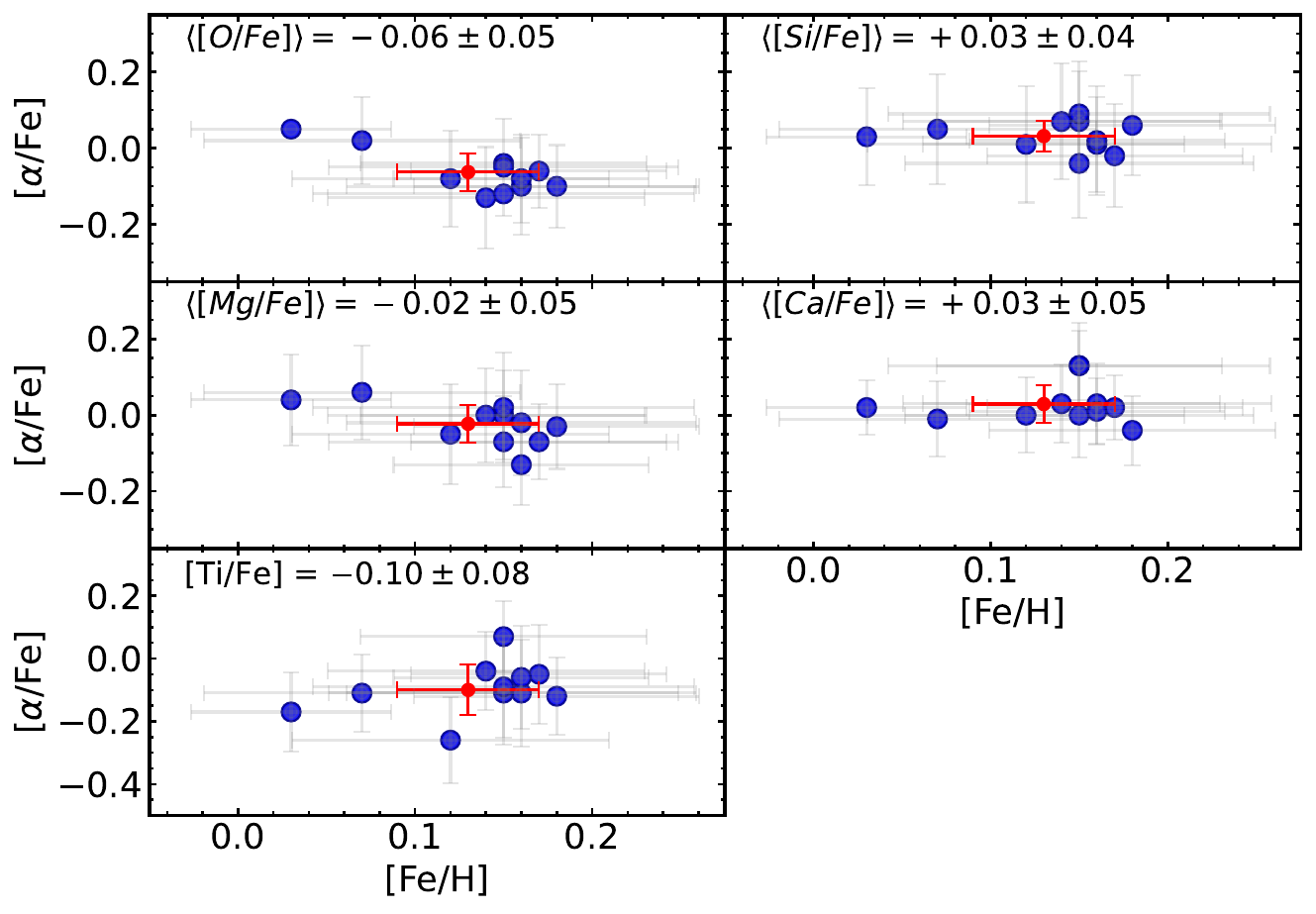}
\caption{[$\alpha$/Fe] vs. [Fe/H] diagram for NGC 6705 stars. In each panel, the red circle indicates the mean abundance ratios of O, Mg, Si and Ca, and Ti as a function of mean metallicity, with the error bars representing the standard deviation.}
\label{fig:alfaFe_6705}
\end{figure*}

The main goal of this study was to do a detailed spectroscopic analysis of APOGEE spectra of NGC 6705 red-giants and, in particular, probe their $\alpha$-element abundances. Figure \ref{fig:alfaFe_6705} summarizes our results in the form of the [$\alpha/\rm Fe$] versus [Fe/H] diagram for the five $\alpha$-elements studied: O, Mg, Si, Ca, and Ti. Each panel depicts the mean [$\alpha$/Fe] (represented by the red circles) and their corresponding standard deviations. The mean abundances obtained for the eleven studied red-giants are shown in each panel.
On the solar scale, our sample exhibits on average a titanium-to-iron ratio ($\rm \langle$ [Ti/Fe]$\rangle$ = $-$0.10 $\pm$ 0.08), and an oxygen-to-iron ratio slightly below solar ($\rm \langle [O/Fe]\rangle=-0.06\pm0.05$), while the mean calcium and silicon abundance ratios are marginally higher than solar scaled values, but not significantly so ($\rm \langle[Ca/Fe]\rangle=+0.03\pm0.05$; $\rm \langle [Si/Fe]\rangle=+0.03\pm0.04$). 
The magnesium abundance for the cluster is also not enhanced and has a mean value slightly below solar ($\rm \langle [Mg/Fe]\rangle=-0.02\pm0.05$). All in all, our analysis does not find NGC 6705 to be $\alpha$-enhanced and this is corroborated by the average of the four $\alpha$-elements studied, which yields $\langle [\alpha/\rm Fe]\rangle =$ $-$0.03 $\pm$ 0.05.

\begin{figure*}
\centering
\includegraphics[width=0.7\textwidth]{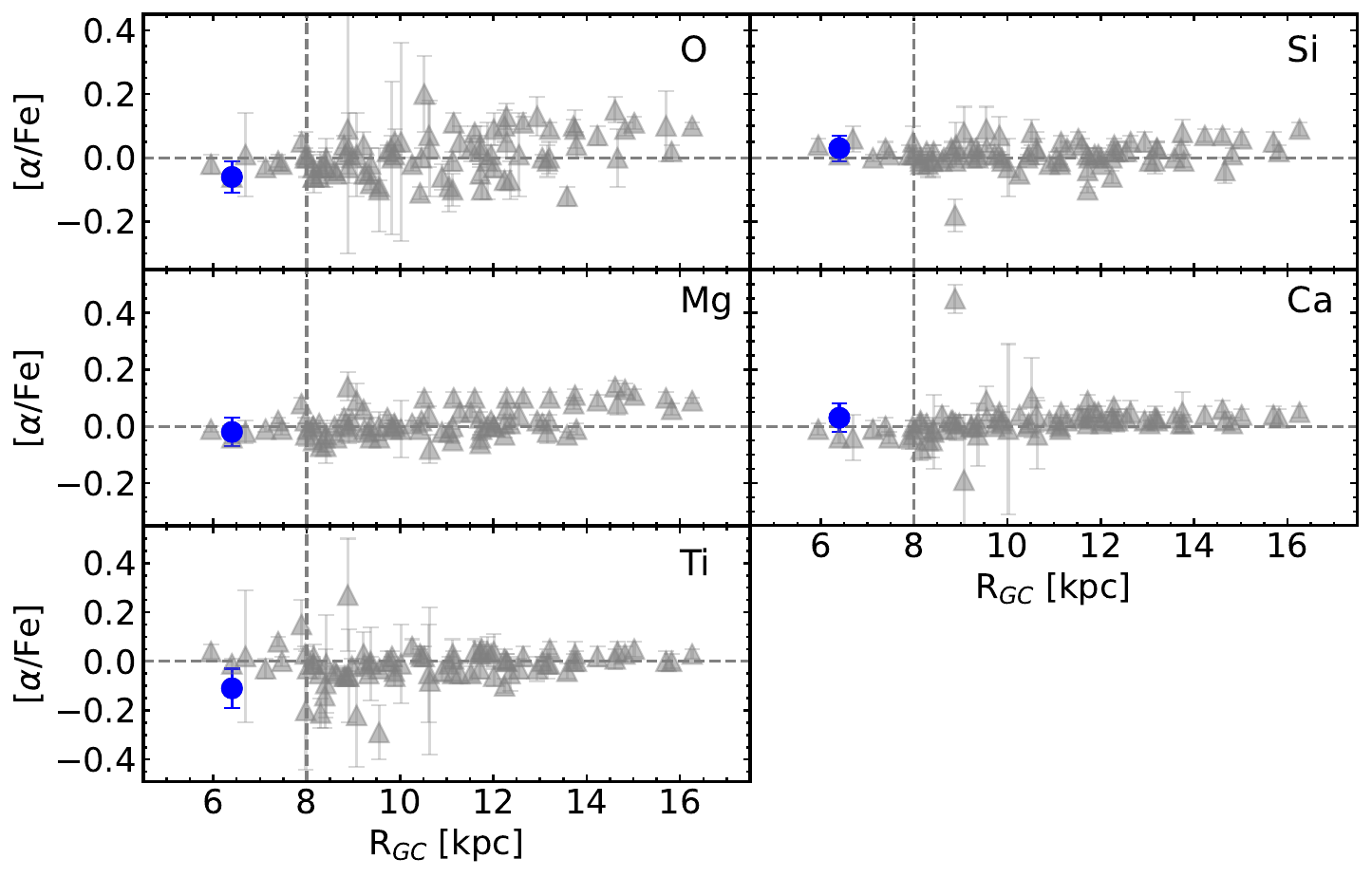}
\caption{The plot shows the cluster mean abundances of [$\alpha$/Fe] as a function of the cluster galactocentric distance (x-axis). The gray triangles represent the OCCAM data from Myers et al. \citeyear{myers2022AJ....164...85M}, while blue circles represent the results of this work.}
\label{fig:trend_rgc}
\end{figure*}

Given the location of NGC 6705 in the inner Galactic disk, it is also of interest to investigate whether the measured [$\alpha$/Fe] abundances for this open cluster are consistent with those of other open clusters residing in the inner galaxy, and how their $\alpha$-element abundance pattern contrasts with the galactic abundance gradients. 
The mean Fe abundances obtained here for NGC 6705 is in line with the gradients for [Fe/H] versus R$_{GC}$ presented in Figure 4 of \cite{myers2022AJ....164...85M}.
The five panels of Figure \ref{fig:trend_rgc} show the O, Mg, Si, Ca, and Ti over Fe ratios as a function of galactocentric distance (R$_{GC}$) for the open clusters from the OCCAM sample \citep[][gray triangles]{myers2022AJ....164...85M}; our results for NGC 6705 are represented by the solid blue circles. 
The abundances in \cite{myers2022AJ....164...85M} are calibrated abundances from DR17 and these were computed using the plane parallel radiative transfer code Synspec and in non-LTE for the elements Mg, Si and Ca \citep{osorio2020AA...637A..80O}.
A simple inspection of Figure \ref{fig:trend_rgc} shows that the ratios for the five $\alpha$-elements in NGC 6705 are overall consistent with and do not fall above the results for the other open cluster in the OCCAM sample and these seem to be in line also with the general trend of metallicity with R$_{GC}$ for the inner galaxy, with all open clusters in the inner galaxy having [$\alpha$/Fe] ratios roughly around the solar value. 

\subsection{Na, Al, K, V, Cr, Mn, Co, and Ce Abundance Patterns \label{sec:ce}}

\begin{figure*}
\centering
\includegraphics[width=1.2\textwidth]{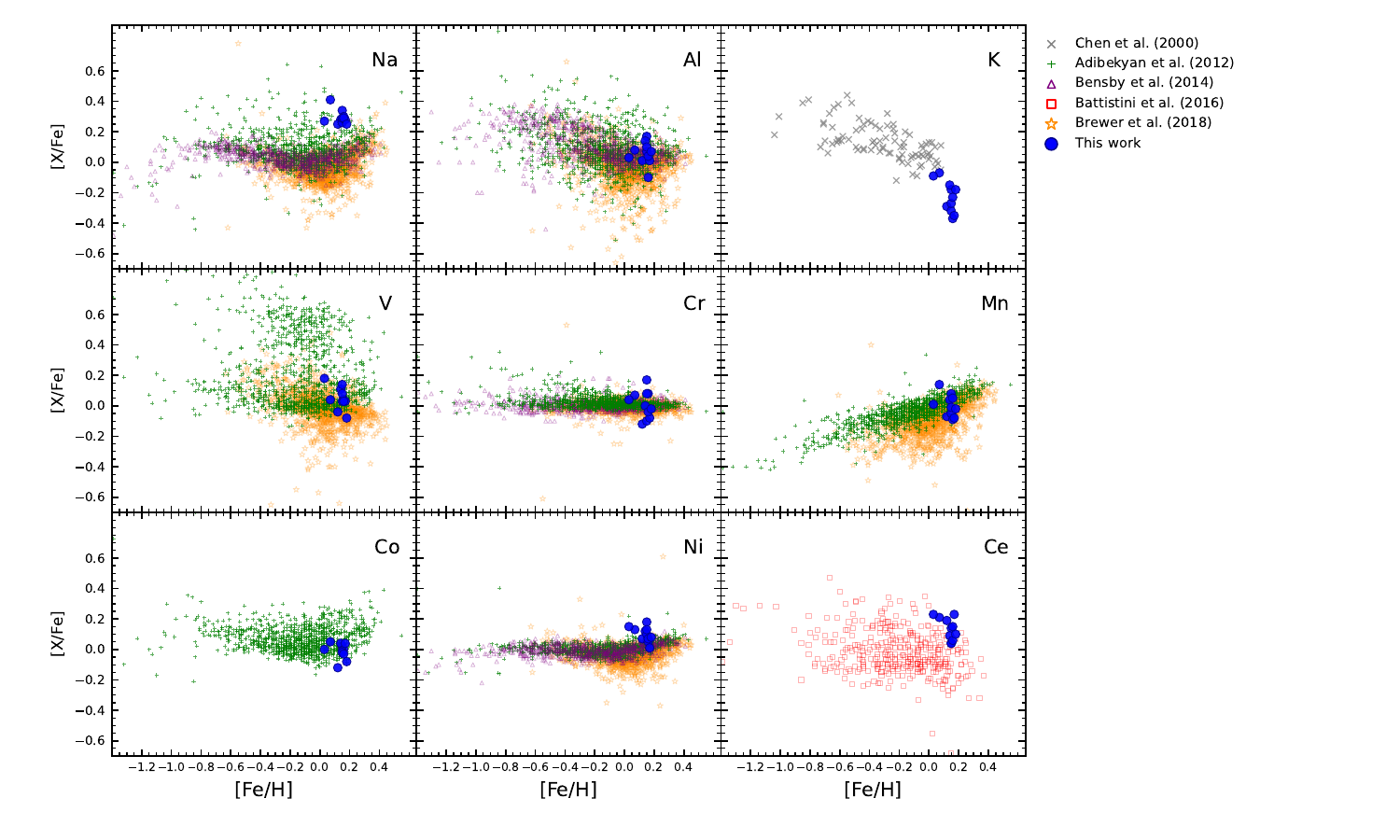}
\caption{Galactic trends of [X/Fe] as a function of [Fe/H] for the stars in the open cluster NGC 6705 (blue circles). Field dwarf stars in the thin and thick disk are from \citet[][gray xs]{chen2000A&AS..141..491C}, \citet[][green pluses]{adibekyan2012A&A...545A..32A}, \citet[][purple triangles]{bensby2014A&A...562A..71B}, \citet[][red squares]{battistini2016A&A...586A..49B}, and \citet[][orange stars]{brewer2018ApJS..237...38B}.}
\label{fig:XFe_6705}
\end{figure*}

The [X/Fe] versus [Fe/H] ratios for the other studied species besides CNO and $\alpha$-elements are presented as filled blue circles in Figure \ref{fig:XFe_6705}, along with abundance results for Galactic field stars from five high-resolution optical studies in the literature \citep{chen2000A&AS..141..491C,adibekyan2012A&A...545A..32A,bensby2014A&A...562A..71B,battistini2016A&A...586A..49B,brewer2018ApJS..237...38B}. The three top panels of Figure \ref{fig:XFe_6705} show the odd-Z elements Na, Al and K. The derived Al abundances fall within the field star distribution at $\rm [Fe/H]\sim +0.15$, while for sodium the abundances of NGC 6705 red-giants fall above the trend; this likely is a mixing signature, as discussed in Section \ref{sec:mixing}. 
For K, the metallicity range probed for the field stars \citep{chen2000A&AS..141..491C} stops at around solar [Fe/H] and the [K/Fe] values for NGC 6705 simply extend the downwards trend of [K/Fe] versus [Fe/H]. For the Fe-peak elements V, Cr, Mn, and Co, our results also generally fall within the field star trends, noting, however, that there is more scatter in our Cr abundances when compared with the very thin (and flat) trend for the field stars obtained from the optical studies. For Ni, the derived abundances for some of the stars fall above the field star trend, the latter being also quite well defined according to the results in \cite{bensby2014A&A...562A..71B} and \cite{adibekyan2012A&A...545A..32A}.

The only heavy element analyzed from APOGEE spectra is the s-process element Ce, which is produced mainly in AGB stars \citep{cescutti2022Univ....8..173C}. The [Ce/Fe] vs [Fe/H] abundance pattern for the field stars from \cite{battistini2016A&A...586A..49B}, shown as comparisons in Figure \ref{fig:XFe_6705}, overall exhibit some scatter. The mean Ce abundance obtained for NGC 6705 red giants is enhanced, with $\rm \langle [Ce/Fe]\rangle=+0.13 \pm 0.07$, falling in the upper envelope of the abundance distribution of field stars at [Fe/H] $>$ 0.0. In addition, the observed [Ce/Fe] enhancement for NGC 6705 is in-line with previous findings that [Ce/Fe] ratios are a function of age \citep{casali2020A&A...639A.127C}. The observed chemical pattern resembles the enhancements in [Ce/Fe] observed in other young open clusters, which is larger than typical values of [Ce/Fe] in old open clusters with similar metallicities \citep[see Figure 6 in][]{sales2022ApJ...926..154S}. 

\section{Conclusions} \label{sec:conclusions}
NGC 6705 is a young open cluster that serves as an abundance benchmark for the inner Milky Way young stellar populations ($\sim 1-5\times10^{8}$ yrs). 
The red giant members of this cluster are also good stellar samples in which to probe evolution along the RGB and RC in 3$-$4 M$_{\odot}$ metal-rich giants.

Previous works in the literature found this benchmark young open cluster to be $\alpha$-enhanced; a puzzling result as young stars are expected to be formed from material already enriched in Fe from SN Ia, which would result in low values of [$\alpha$/Fe]. The abundance patterns for NGC 6705 were discussed by \cite{Casamiquela2018AA...610A..66C} in the context of the population of young and $\alpha$-enhanced field stars found from CoRoT and Kepler data \citep{Chiappini2015, Martig2015,queiroz2023AA...673A.155Q}.

The population of young-$\alpha$-enhanced stars identified in the Galaxy, however, has now been shown to be old and to possibly result from binary mergers or mass transfer as suggested by 
\citet{izzard2018MNRAS.473.2984I,hekker2019MNRAS.487.4343H}. This scenario was strengthened by the results of \cite{miglio2021AA...645A..85M} from their analysis of old, thick-disk giants in the Kepler field that have been found to be overmassive, meaning that their initial birth masses were increased by mass transfer or mergers while evolving up the RGB. 

The possibility that the young cluster NGC 6705 is $\alpha$-enhanced was further tested here based on the derived abundances of five $\alpha$-elements (O, Mg, Si, and Ca, and Ti), finding $\rm \langle [\alpha/Fe]\rangle = -0.025 \pm 0.051$), and indicating that NGC 6705 does not exhibit $\alpha$-enhancement. Our results are consistent with the expectation that the young open cluster NGC 6705 in the Galactic disk has solar-like values of the [$\alpha$/Fe] ratio. 

This study presented a quantitative spectroscopic analysis of eleven red giant stars members of the open cluster NGC 6705, determining abundances of the elements C, N, Na, Al, K, Ti, V, Cr, Mn, Fe, Co, Ni, and Ce, as well as the $\alpha$-elements O, Mg, Si and Ca. The analysis was carried out in LTE, using MARCS spherical model atmospheres with the spherical radiative transfer program Turbospectrum. The mean abundances obtained for the cluster are presented in Table \ref{tab:mean_abu}.

Our results from the analysis of APOGEE NIR spectra for the NGC 6705 stars find an average metallicity of $\rm \langle [Fe/H]\rangle=+0.13\pm 0.04$ dex, which is in agreement with the general trend of increasing metallicity with decreasing distance from the Galactic center. The ratios of the Fe-peak elements are found to track Galactic trends, as defined by field stars, with the mean values of [X/Fe] for V, Cr, Mn, Co, and Ni being +0.05, +0.01, 0.00, $-0.01$, and +0.10, respectively. The [Al/Fe] abundance ratios also fall within the distribution of field stars, while the [K/Fe] values appear to extend the downward trend of [K/Fe] vs. [Fe/H]. 
Results for the s-process element cerium, with $\rm \langle [Ce/Fe]\rangle = +0.13$, are similar to the Ce abundances observed in other young open clusters with similar metallicities to NGC 6705 and consistent with an increase in the [Ce/Fe] ratio with decreasing age \citep{sales2022ApJ...926..154S}.

The red-giant members of NGC 6705 exhibit the low-$^{12}$C and enhanced-$^{14}$N abundance signature of 1st dredge-up on the RGB, with mean values of $\rm [^{12}C/Fe]=-0.16$ and $\rm [^{14}N/Fe]=+0.51$. 
These abundances are in quantitative agreement with 3 M$_{\odot}$ $-$ 4.0 M$_{\odot}$ stellar models from \cite{lagarde2012AA...543A.108L}.
The sample here contains both candidate RGB (7) and RC (4) stars and a comparison of the $^{12}$C and $^{14}$N abundances between the two groups reveals no significant differences, indicating no measurable ``extra mixing'' processes as these metal-rich 3.3 M$_{\odot}$ stars evolve up the RGB and then onto the He-burning red clump. 
In addition to carbon-12 and nitrogen-14, oxygen, sodium, and aluminum abundances were compared to stellar models in order to test for deep mixing signatures.  Sodium was found to be enhanced significantly, with $\rm [Na/Fe]=+0.29\pm0.04$, in general agreement with stellar evolution model predictions from \cite{lagarde2012AA...543A.108L} and in-line with what was previously concluded in \cite{smiljanic2016AA...589A.115S}. For aluminum, however, we find non-enhanced values of [Al/Fe] for NGC 6705, in contrast with what was found in the latter study.
The values of [O/Fe] and [Al/Fe] for NGC 6705 were found to be roughly solar, within small uncertainties and a comparison to models by \cite{lagarde2012AA...543A.108L} also finds agreement with our O and Al abundance results for stellar masses of 3.3 M$_{\odot}$. 
Summarizing, at the masses of the NGC 6705 red giants, standard stellar evolution models agree well with the observationally-derived abundances of $^{12}$C, $^{14}$N, $^{16}$O, Na, and Al.

\section*{Acknowledgements}
We thank the referee for suggestions that improved the paper. V. L-T. acknowledges the financial support from Coordena\c{c}\~{a}o de Aperfei\c{c}oamento de Pessoal de Nível Superior (CAPES).
K.C. acknowledges that her work is supported, in part, by the National Science Foundation through NSF grant No. AST-2206543.
D.S. thanks the National Council for Scientific and Technological Development – CNPq.
\addcontentsline{toc}{section}{Acknowledgements}

Funding for the Sloan Digital Sky Survey IV has been provided by the Alfred P. Sloan Foundation, the U.S. Department of Energy Office of Science, and the Participating Institutions. SDSS-IV acknowledges support and resources from the Center for High-Performance Computing at the University of Utah. The SDSS website is www.sdss.org.
SDSS-IV is managed by the Astrophysical Research consortium for the Participating Institutions of the SDSS Collaboration including the Brazilian Participation Group, the Carnegie Institution for Science, Carnegie Mellon University, the Chilean Participation Group, the French Participation Group, Harvard-Smithsonian Center for Astrophysics, Instituto de Astrof\'isica de Canarias, The Johns Hopkins University, 
Kavli Institute for the Physics and Mathematics of the Universe (IPMU) /  University of Tokyo, Lawrence Berkeley National Laboratory, Leibniz Institut f\"ur Astrophysik Potsdam (AIP),  Max-Planck-Institut f\"ur Astronomie (MPIA Heidelberg), Max-Planck-Institut f\"ur Astrophysik (MPA Garching), Max-Planck-Institut f\"ur Extraterrestrische Physik (MPE), National Astronomical Observatory of China, New Mexico State University, New York University, University of Notre Dame, Observat\'orio Nacional / MCTI, The Ohio State University, Pennsylvania State University, Shanghai Astronomical Observatory, United Kingdom Participation Group,
Universidad Nacional Aut\'onoma de M\'exico, University of Arizona, University of Colorado Boulder, University of Oxford, University of Portsmouth, University of Utah, University of Virginia, University of Washington, University of Wisconsin, Vanderbilt University, and Yale University.

{\it Facilities: {Sloan}, {Gaia}}.

Software: BACCHUS (Masseron et al. \citeyear{masseron2016ascl.soft05004M}), Turbospectrum (Alvarez \& Plez \citeyear{alvarez1998A&A...330.1109A}; Plez \citeyear{plez2012ascl.soft05004P}; \href{https://github.com/bertrandplez/Turbospectrum2019}{https://github.com/bertrandplez/Turbospectrum2019}).

\subsection*{Data availability}
The data underlying this article are available in the SDSS Data Release 17 for APOGEE.   These data can be accessed at the following links: \newline
\href{https://www.sdss.org/dr17/irspec/}{https://www.sdss.org/dr17/irspec/}(APOGEE).


\bibliography{references}{}
\bibliographystyle{mnras}

\bsp
\label{lastpage}
\end{document}